\newcommand{\A} {\hat{A}}
\newcommand{\As} {\hat{A}_s}
\newcommand{\Aout}  {\hat{A}_s^{out} }
\newcommand{\Aoutdag} {\hat{A}_s^{\dagger \mspace{1.0mu} out } }
\newcommand{\Ain}{\hat A_s^{in} }
\newcommand{\Aindag}{\hat A_s^{{ \dagger \mspace{1.0mu} in }} }
\newcommand{\ah} {\hat{a}}
\newcommand{\as} {\hat{a}_s}
\newcommand{\app} {\hat{a}_p}

\newcommand{\ain}{\hat a_s^{{in}} }
\newcommand{\aindag}{\hat a_s^{{ \dagger\mspace{1.0mu} in}} }

\newcommand{\apc}{{\alpha}_p}

\newcommand{\rhop}{ \rho_p}  %modulo del profilo della pompa 

\newcommand {\Gone } {G^{(1)} }
\newcommand {\chidue } {\chi^{(2)} }

\newcommand{\Uj}{ {\mathcal K}_j}
\newcommand{\Uu}{ \mathcal K_1}
\newcommand{\Vv}{ \mathcal K_2}
\newcommand{\f}{ f}
\newcommand{\F}{  F}
\newcommand {\Fcorr}{F_{corr}} 
\newcommand {\Fcoh}{F_{coh}} 
\newcommand {\mucorr}{\mu_{corr}} 
\newcommand {\mucoh}{\mu_{coh}} 

\newcommand{\DD}{\mathcal D}                   %  Phase mismatch
\newcommand{\D} { D}                                      % Plane-wave  Phase mismatch
\newcommand{\Db} { \bar D}                               % Plane-wave  Phase mismatch x crystal length 
\newcommand{\Dbz} { \bar D_0}                               % Plane-wave  Phase mismatch x crystal length  in collinear degenerate point 
\newcommand{\taugvm}{\tau_{\textsc{gvm}}  }
\newcommand{\lwoff}{l_{\mathrm{woff}}  }
\newcommand{\Dxisp}{\vec{ \xi}_\textsc{wo}  }

\newcommand{\Omgvd}{\Omega_{\textsc{gvd}}  }
\newcommand{\taugvd}{\tau_{\textsc{gvd}}  }
\newcommand{\qdiff}{q_{\mathrm{diff}}  }

\newcommand{\w}{\vec{w}}			% 3-D q,omega

\newcommand{\wpr} {\vec{w}\mspace{2.0mu}'}%{\vec{w}{\mspace{1.0mu}^\prime\mspace{-2.0mu}}}
\newcommand{\vxi}{\vec{\xi \mspace{1.0mu}}}	
\newcommand{\vxipr}{\vec{\xi}{\mspace{1.0mu}^\prime\mspace{-2.0mu}}}
	
\newcommand{\Om}{\Omega}
\newcommand{\q}{\vec{q}}

\newcommand{\rr}{\vec{r}}
\newcommand{\llangle} { \left \langle}   	
\newcommand{\rrangle} { \right\rangle}   	
\newcommand{\ex} {\vec{e}_x}
\newcommand{\ey} {\vec{e}_y}

\newcommand{\sinc}{{\rm sinc}}

\newcommand{\nn}{\nonumber}
\newcommand{\bsub}{\begin{subequations}}
\newcommand{\esub}{\end{subequations}}
\newcommand{\beq}{\begin{equation}}
\newcommand{\eeq}{\end{equation}}
\newcommand{\beqa}{\begin{align}}
\newcommand{\eeqa}{\end{align}}
\documentclass[noinfo]{rstransa}%%%%where rstrans is the template name
\usepackage{amsmath} 
\usepackage{empheq}
\usepackage{amssymb}
\usepackage{graphicx}
\usepackage{float}   
\usepackage{braket}
\usepackage{xcolor}
\DeclareMathSymbol{\Rho}{\mathalpha}{operators}{"50}

\titlehead{Research}

\begin{document}

%%%% Article title to be placed here
\title{Unified space-time description of pulsed twin beams} %A quasi-stationary model for pulsed twin beams}
\author{%%%% Author details
Alessandra Gatti $^{1}$, Enrico Brambilla$^{2}$ and Ottavia Jedrkiewicz$^{1,2}$}
%%%%%%%%% Insert author address here
\address{$^{1}$Istituto di Fotonica e Nanotecnologie del CNR, Piazza Leonardo da Vinci 32, 20133 Milano, Italy.\\
$^{2}$Dipartimento di Scienze e Alta Tecnologia, Universit\`a dell'Insubria, Via Valleggio 11, 22100 Como,Italy}

%%%% Subject entries to be placed here %%%%
\subject{xxxxx, xxxxx, xxxx}
%%%% Keyword entries to be placed here %%%%

%%%% Insert corresponding author and its email address}
\corres{Alessandra Gatti\\
\email{Alessandra.Gatti@ifn.cnr.it}}

%%%% Abstract text to be placed here %%%%%%%%%%%%
\begin{abstract}
This work provides a mathematical derivation of a quasi-stationary model for multimode  parametric down-conversion, which was presented in [Gatti {\it et al.}, {\it Sci. Rep.} 13, 16786] with heuristic arguments. The model is here derived from the 3D+1 propagation equation of the quantum fields in a nonlinear crystal, and its approximations discussed thoroughly.  Thanks to its relative simplicity, and to the fact that it is valid in  any gain regime, both at a  quantum and classical level, it allows a unified  description of  disparate experimental observations conducted over the last 20 years, often described in the past  by means of limited ad hoc models. 
%In this work we present a simple semi-analytical quasi stationary model describing, through simple formulas, all the the features of multimode Parametric Down-Conversion processes in any gain regime, both at a classical and quantum level. We derive the model in a rigorous way from the quantum propagation equations of light in a nonlinear medium. We also explore some of its predictions by reviewing the experimental observations that can be found in the literature relative to the high-gain regime.
\end{abstract}
%%%%%%%%%%%%%%%%%%%%%%%%%%%
%%%%%%%%%% Insert the texts which can accomdate on firstpage in the tag "fmtext" %%%%%
\begin{fmtext}
\section*{Introduction}
During propagation  in  certain nonlinear materials, %(nonlinear $\chidue$ crystals),  
light photons from an intese optical beam
can occasionally split into pairs of  lower-energy photons, traditionally  named ``twins''  because of their simultaneity \cite{Burnham1970}.  
Twin photons share not only the time of their birth, but essentially all their physical properties.  In  a quantum mechanical description, this implies that their state is nonseparable ({\it entangled})  with respect to all light degrees of freedom.
Moreover, since the process is intrinsically broadband, twin photons can be generated in a large number of independent space-time modes \cite{Law2000,Eberly2004, Gatti2012}. 
These features made  parametric down-conversion  (PDC),  as classically is called  the  process,  a favorite choice  for  quantum imaging  (see the reviews \cite{Gatti2008,Genovese2016,Padgett2019}), which exploits their spatial transverse entanglement,   or entangled two-photon microscopy \cite{Tabakaev2021,Raymer2021b} and spectroscopy \cite{Schlawin2018},  which relies on their time-frequency correlation.  However, the same features make the  theoretical description of multimode PDC  challenging. A well-known analytical model 
exists in the low-gain perturbative limit \cite{Ghosh1986}, see 
%%% Break della pagina singola colonna, per spostarlo copiare tutto il terzetto di comandi e spostarlo 
\end{fmtext}
\maketitle
\noindent 
also \cite{Atature2002,Gatti2009}, in which photon pairs are individually generated and detected.  
Though, real-world
 implementations of quantum imaging and microscopy are likely to rather require   the bright entangled beams (or the {\it bright squeezed vacuum}\cite{Sharapova2015}) generated at high-gain. In this regime,  an analytical model  is available 
 only for a plane-wave pump \cite{Klyshko1988,Kolobov1999, Brambilla2001, Brambilla2004,Gatti2003}, which, besides being nonphysical,  does  not correspond to modern setups based on the use of short laser pulses. 
Several  theoretical approaches were developed to account for the finite size of the laser pump, including  numerical simulations of stochastic equations \cite{Brambilla2004}, generalized Bogoliubov relations in the space-time domain \cite{Brambilla2004}  or in the time-domain only \cite{Christ2013} (see also \cite{Kulkarni2022} for a classical version of these equations), the integro-differential equations described in \cite{Sharapova2020}, and various kind of Schmidt-mode decompositions \cite{Wasilewski2006,Sharapova2015}. However, all these approaches, despite being potentially very accurate,
 still need important numerical treatments to extrapolate results. \\
In a recent work of ours \cite{Gatti2023}  a substantially simpler and  semi-analytic  model was presented  on the basis of heuristic arguments. 
%that we named {\em quasi stationary } model. 
In the present work, we derive the same {\it quasi stationary }  model (or different form of the same model) 
 in a more rigorous way from the quantum propagation equations of light in a nonlinear medium, and we discuss thoroughly the assumptions on which it is based. In order not to burden the formalism, we focus on the the widely used  type I quasi-degenerate PDC, in which twin photons are generated with the same polarization and close to the same frequency. Generalizations to other configurations are not difficult, but will be discussed elsewhere. 
 In the second part of the work, we explore some of its predictions, in the framework of different experimental observations performed in the last 20 years.  
 Although 
the model describes any gain regime of PDC, we focus on the high-gain, in which many disparate experimental observations, previously described by {\it ad hoc} models, often limited to  time-only  or space-only domains,  can find a valid description within our model. 
%%%%%%%%%%%%%%% End of first page %%%%%%%%%%%%%%%%%%%%%
\section{Quasi-stationary model for pulsed PDC}
\subsection{Background and definitions}
Our analysis starts from the equations  that describe the propagation in a nonlinear $\chidue$ medium of the quantum fields  associated with an input  pump beam, of central frequency $\omega_p$,  and with the down-converted signal   of central frequency $\omega_s={\omega_p}/{2}$.  We assume paraxial propagation  around a mean direction $z$ (Fig.\ref{fig_scheme}), and introduce the operators: 
\begin{align}
\A_j(\rr, t,z)   &=  \int \frac{d^2 \q}{2\pi}   \int \frac{d\Om}{\sqrt{2\pi}}  e^{i \q \cdot \rr } e^{ - i \Om  t }   \A_j (\q, \Om, z)   \qquad j=s,p 
\label{AFourier}
\end{align}
where  $\rr=x\ex +y \ey$  is the  position in the transverse plane,  $\q= q_x \ex + q_y \ey $  is the transverse  wave-vector,   $\Om$ is the  frequency offset from the carriers,    and   dimensions are such that 
 $ \A_j^\dagger (\rr,t,z)  \A_j  (\rr,t,z)$  is  a photon number per unit area and time.  
Their evolution along the slab is best described in an interaction picture in which the linear propagation in the medium, accounting for diffraction and dispersion at any order, is subtracted  (see \cite{Gatti2003,Brambilla2012} for details), by setting
\beq
  \hat A_j  (\q,\Om,z) = e^{ i k_{z j} (\q,\Om) z}  \hat a_j (\q, \Om,z) ,  \qquad \text{where} \quad k_{jz} (\q, \Om) = \sqrt {k_j^2(\q,\Om ) -q^2}  \, , 
\label{ajdef}
\eeq
$ k_j(\q, \Om)$  being
%= n_j (\q, \omega_j + \Om )\frac{ \omega_j + \Om}{c}$ 
 the wave-number of the j-th wave (it depends on the direction of propagation through $\q$ only for the extraordinary wave).  
Then,  the lowercase operators $\ah_j$ evolve slowly under the action of the nonlinear interaction.\footnote{Notice that for broadband PDC this approximation is far better than the usual slowly varying envelope approximation \cite{LPBbook2015}} %which subtracts only  the  oscillation at the fixed carriers}
In common  implementations  of PDC aimed at generating quantum states of light, the parametric gain is not too large, which justifies  the {undepleted pump} approximation, in which %the  pump beam   is  unaffected by down-conversion. Then, 
$\app (\q,\Om, z) = \app (\q,\Om, 0)  $ and the pump operator is substituted by a c-number field. 
 By adopting a   shorthand notation,  
 in which  
 \begin{align}
 & \vxi:  = (x,y, t),  \qquad & \w : = (q_x,q_y, \Om), 
  \end{align}
 with the convention for the scalar product: 
 $
  \w\cdot \vxi = x q_x +y q_y - \Om t ,
 $
% are the  space-time  vector and  the conjugate Fourier vector, respectively,
% with the convention for the scalar product $\w\cdot \vxi = \q\cdot \rr  - \Om t $, 
the evolution of the signal field along the slab is  described by the linear equation (see \cite{Gatti2003,Brambilla2004} for a derivation): 
\begin{align} 
\frac{\partial  \as }{\partial z}   (\w, z )   &=  \frac{g}{l_c}  \int 
 \frac{d^3 \w_0 }  {(2\pi)^{\frac{3}{2}} }  \, \apc (\w_0 ) \, 
 \as^\dagger(\w_0-\w, z)  e^{-i \DD (\w;  \w_0-\w) z }  ,
\label{prop}
\end{align}
where: $\apc (\w) = \int  \frac{d^3 \vxi }  {(2\pi)^{{3}/{2}} }   e^{-i \w\cdot \vxi} \apc (\vxi)$ is the Fourier profile of the input pump field, normalized so that $\apc (\vxi =0)=1$; 
  $g$ is the  dimensionless gain parameter, proportional to the nonlinear susceptibility, the crystal length and the pump peak amplitude; 
\begin{align}
\DD ( \w;   \w_0 - \w)  &:=  k_{sz} (\w) +  k_{sz} (\w_0 - \w )  - k_{pz} (\w_0) 
\label{PM}
\end{align}
%is  the phase mismatch of   the  down-conversion process in which  a pump photon in mode $\w_0$ splits  into a pair of signal and idler photons in modes $\w$ and $\w_0 -\w$, as allowed by energy -transverse momentum conservation. 
is the phase mismatch of  the elementary  process  in which a  pump photon in mode  $\w_0$  splits into    a photon pair  in modes $\w$ and $\w_0 -\w$, with conservation of the  energy  and transverse momentum. 
\subsection{Quasi-stationary approximation} 
%%%figura schema propagazione con spettro %%%%%%%
\begin{figure}[t]
\centering
{{\includegraphics[width=0.82\linewidth]{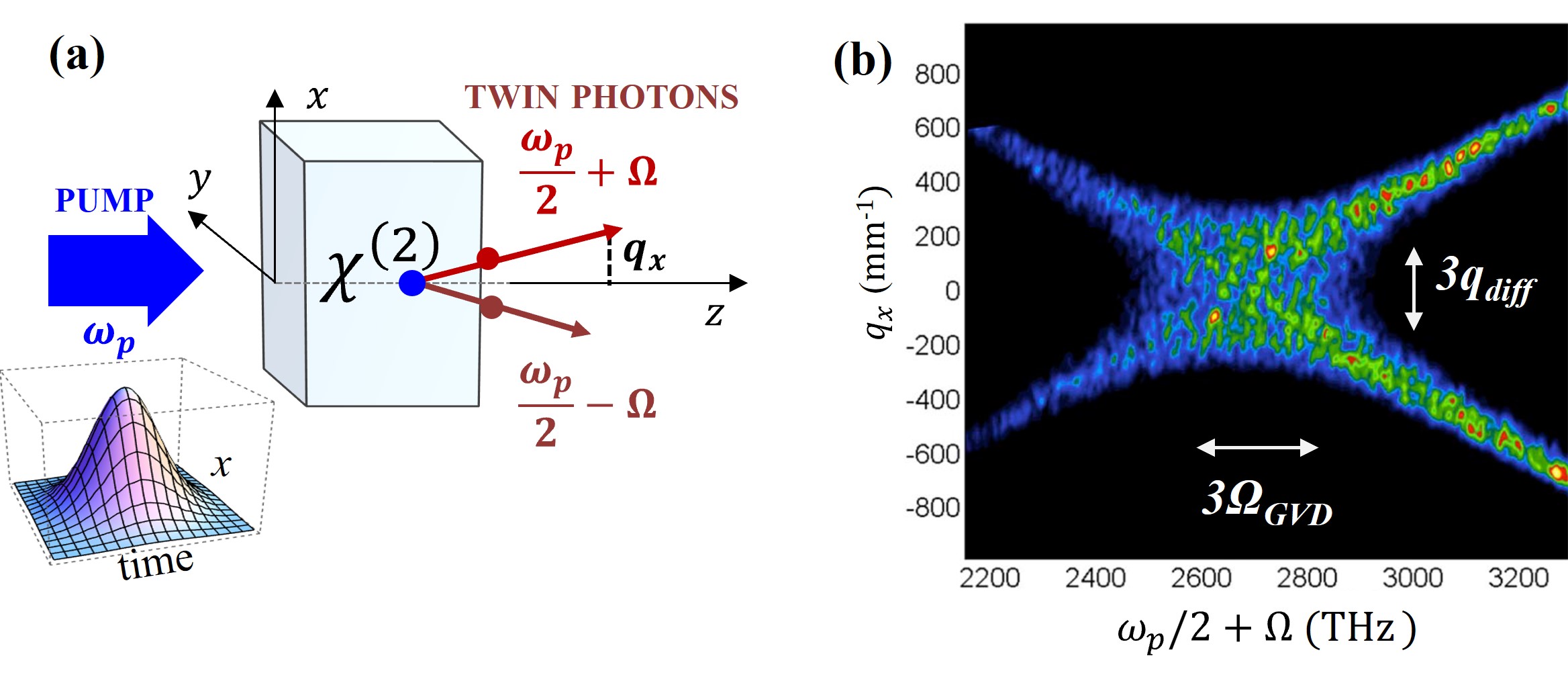}}}
\caption{(a)  Twin-photon generation in a $\chidue$ medium. (b) Example of Fourier spectrum, measured in Ref.\cite{Jedr2007}, generated from a 2 mm $\beta$–Barium borate (BBO) crystal pumped at $352$ nm by a 1 ps,  $ 200\, \mu$m  FHWM pulsed beam. For these parameters the bandwidths,  defined in Eq.\eqref{PMbandwidth},  are 
$\Omgvd=74  \text{ THz}$, $\qdiff=86 \text{ mm}^{-1}$. }
\label{fig_scheme}
\end{figure}
%%%%
 The solution  of equation \eqref{prop} is a  generalized Bogoljubov  transformation, similar to that studied in \cite{Brambilla2004}, linking in a nonlocal way operators at  position $z$  to those at the input $\ain (\w):= \as (\w, 0)$:
 \begin{align}
\as (\w,z) = \int d^3\wpr \, \left[ \Uu (\w, \wpr,z) \ain (\wpr) + \Vv  (\w, \wpr, z) \aindag (-\wpr) \right] \, .
\label{Bogu}
\end{align}
where the kernels   of the transformation 
%satisfy the unitarity relations \nota{Mettero' alòcune equazioni nei supplementari}: 
%\begin{align}
% &\int d^3 \w \, \left[\Uu (\w_1, \w ) \Uu^* (\w_2, \w) - \Vv (\w_1, \w ) \Vv^* (\w_2, \w)  \right]= \delta (\w_1-\w_2)\\
% & \int d^3 \w\,  \left[\Uu (\w_1, \w) \Vv (\w_2,- \w) - \Vv (\w_1, \w ) \Uu (\w_2, -\w)  \right]=0
%\end{align}
%and 
obey the evolution equations
\beq
\begin{aligned}
\frac{\partial  \Uu }{\partial z}  (\w,\wpr ) &= \frac{g}{l_c}  \int  \frac{d^3 \w_0 }  {(2\pi)^{\frac{3}{2}} }  \, \apc (\w_0 ) \, 
 \Vv^*(\w_0-\w, -\wpr)  e^{-i \DD (\w;  \w_0-\w) z }  \\
 \frac{\partial  \Vv  }{\partial z} (\w,\wpr ) &= \frac{g}{l_c}  \int  \frac{d^3 \w_0 }  {(2\pi)^{\frac{3}{2}} }  \, \apc (\w_0 ) \, 
 \Uu^*(\w_0-\w, -\wpr)  e^{-i \DD (\w;  \w_0-\w) z }  \label{Kprop}
\end{aligned}
\eeq
with initial conditions $\Uu (\w,\wpr)|_{z=0}=\delta(\w-\wpr) $ and $ \Vv (\w,\wpr)|_{z=0}=0$. 
\par 
Up to now we  merely reformulated the propagation equation \eqref{prop}.  
A well-know solution exists in the limit of a  monochromatic plane-wave  pump (PWP) \cite{Klyshko1988,Kolobov1999,Brambilla2001, Brambilla2004,Gatti2003}, $  \apc (\w_0 ) \to    (2\pi)^{{3}/{2} }   \delta(\w_0)$, in which the kernels reduce to Dirac-delta functions.  %and the Bogulobov relation \eqref{Bogu} becomes {\em local $ in the Fourier space, coupling only phase conjugate modes $\w \leftrightarrow -\w$ 
The {\it  quasi-stationary} approximation aims at going beyond the unphysical PWP limit,  assuming that  the pump pulse  is finite but its Fourier spectrum is narrow enough  that there exist  two-well separated scales of variation: 
\\
-- A {\em fast} scale, characterizing the decay of the coherence and correlation functions in the Fourier domain. 
An educated guess, based on the low gain perturbative solution, is that this scale is associated with the Fourier spectrum of the pump. Indeed, energy-momentum conservation imposes that  in the spontaneous regime, the energy and transverse momentum of twin photons sum up to the those of the pump beam.  As we shall see,  at high gain this fast scale undergoes some  broadening \cite{Jedr2006,Allevi2014,Gatti2023}. \\
-- A {\em slow} scale, characterizing the decay of the overall Fourier  spectrum of the PDC emission. This is associated with the scale of variation of  the  phase matching between conjugate modes: 
\beq
l_c \DD (\w; -\w)  \approx
% l_c (2k_s-k_p) +k''_s l_c \Omega^2 -\frac{q^2 l_c}{2k_s} =
 l_c (2k_s-k_p) + \frac{\Omega^2}{\Omgvd^2} {\mathrm sgn} (k''_s ) - \frac{q^2}{\qdiff^2}
 \label{TaylorPM}
\eeq
where the Taylor expansion up to second order  of Eq.\eqref{PM}  has been used, and 
\beq
\Omgvd= |k''_s l_c|^{-\frac{1}{2}}, \qquad  \qdiff= \sqrt{k_s/ l_c}
\label{PMbandwidth}
\eeq
 This scale, which reflects the conservation of longitudinal momentum in the parametric scattering,  also undergoes a broadening at high gain\cite{Spasibko2012,Sharapova2020}. \\
Physically, this  configuration  corresponds to  small correlated light  speckles  inside a broader spectral distribution, as  observed in several high-gain experiments \cite{Jedr2004, Jedr2006,Jedr2007,Brida2009,Brida2009bis, Jedr2012a,Allevi2014,Sharapova2020}, and  depicted in the example of Fig.\ref{fig_scheme}b. 
Mathematically, our derivation is based on two assumptions: 
\begin{description}
\vspace{-0.3 cm}
 \setlength{\itemindent}{-2em}
\item[\bf Ansatz 1] As a function of the difference $\w-\wpr$, the kernels $\Uj (\w,\wpr) $ decay on the {\em fast} scale, and are non-zero over a restricted domain $S_0$. As a function of $\w$ (or $\wpr$, or $\frac{\w+\wpr}{2}$) they vary on the {\em slow} scale. As a consequence,  for any $\w_0 \in S_0$  we can approximate 
%$\apc(\w_0) \Uj^*(\w_0-\w, -\wpr) \approx %%\apc(\w_0) \Uj^*(-\w, -\wpr+\w_0) $ at the right hand side of 
$\Uj (\w, \wpr) = \Uj(\w+\w_0,\wpr+\w_0)$ \footnote{In a truly stationary model, this would hold {\em for any} $\w_0$ not just in the fast domain $S_0$}. 
\vspace{0.5em}
\item[\bf Ansatz 2] For any $\w_0 \in S_0$ it holds the approximate identity: 
\beq
\DD(\w,\w_0-\w)= \DD (\w,-\w) - (k'_p -k'_s) \Omega_0 -\frac{\partial k_p}{\partial q_x} q_{0x}
\label{Ansatz2}
\eeq
where we took the x axis in the direction of the walk-off of the Poynting vector ($\frac{\partial k_p}{\partial q_y}=0$). 
\end{description}
\vspace{-0.3 cm}
Notice that while the first ansatz is a very general statement, Ansatz 2 relies on  the specific physical properties of the system. As discussed in Supplementary S.III, it can be obtained by retaining the dominant terms of the  Taylor expansion of the  phase mismatch in the fast variables $\w_0$, and can be nicely satisfied  provided that  $ \Omega_0 \ll  \Omgvd$, $q_0 \ll \qdiff$  (i.e. $ \w_0 \in S_0$) and  that the PDC bandwidth considered is not too large compared to $ \Omgvd$,  see the example or Fig.S3. 
\par 
Introducing the two ansatz in Eq.\eqref{Kprop},  the evolution equations for the kernels become: 
\begin{align}
\frac{\partial   \Uu}{\partial z}  (\w,\w' ) 
%=& \frac{g}{l_c}  \int  \frac{d^3 \w_0 }  {(2\pi)^{\frac{3}{2}} }  \, \apc (\w_0 ) e^{-i \DD (\w; \w_0-\w) z}\Vv^*(-\w, -\wpr-\w_0) \nn \\
 =& \frac{g}{l_c}   e^{-i \D (\w) z} \int  \frac{d^3 \w_0 }  {(2\pi)^{\frac{3}{2}} }  \, \apc (\w_0 ) 
%e^{i \Delta \vxi_{wo}\cdot w_0 \frac{z}{l_c}}\, 
e^{i \left[(k'_p-k'_s) \Omega_0  + \frac{\partial k_p}{\partial q_x} q_{0x}\right] z}
 \Vv^*(-\w, -\wpr-\w_0) 
 \end{align}
 where   $D(\w)$ is a shorthand for $\DD(\w; -\w)$. The equation for $\Vv (\w,\w')$ is obtained by exchanging $ \Uu \leftrightarrow \Vv$. 
These equations become more trasparent if we introduce the Fourier transform of the kernels with respecto to the difference of arguments (fast variable): 
\begin{align}
\Uj (\w,\wpr)= \int  \frac{d^3 \vxi }  {(2\pi)^3} e^{-i (\w -\wpr) \cdot \vxi} \f_j (\w, \vxi)
\label{fdef}
% ; \quad  & F_j (\w,\vxi)= \int  d^3 \w_0 e^{i (\w_0) \cdot \vxi} \Uj (\w, \w - \w_0)
\end{align}
After some not difficult passages,  the equations for the fuctions $\f_j$ are obtained: 
\beq
\begin{aligned}
&\frac{\partial  \f_1 }{\partial z}  (\w,\vxi) =\frac{g}{l_c} \apc\left(\vxi + \Dxisp \frac{z}{l_c}\right) \f_2^* (-\w,\vxi)  e^{-i \D (\w) z}\\
&\frac{\partial  \f_2^* }{\partial z} (-\w,\vxi) =\frac{g}{l_c} \apc^*\left(\vxi + \Dxisp \frac{z}{l_c}\right) \f_1 (\w,\vxi)  e^{i \D (\w) z}
\label{Fprop1}
\end{aligned} 
\eeq
where $\Dxisp =  (-\lwoff, 0, \taugvm) $, with $\lwoff= \frac{\partial k_p}{\partial q_x} l_c$ being the lateral walk-off between the pump and the signal  in crossing the crystal,  and  $\taugvm=l_c( k'_p-k'_s)$ the delay due to group velocity mismatch (GVM). 
%= \frac{l_c}{v_{gp}}- \frac{l_c}{v_{gs}}$, $v_{gj} $ being the group velocities of the two wave-packets. 
 In this way, we  managed to reduce the completely coupled multimode problem \eqref{Kprop} to a system of just two coupled equations for $\f_1  (\w,\xi)$ and $\f_2  (-\w,\xi)$, of much easier solution. Notice, that as long as $\f_j (\w, \xi)$  are not flat functions of the space-time variable $\vxi$, these equations still couple  a large number of PDC modes within the domain $S_0$ through the Bogoljubov tranformation \eqref{Bogu}. 
Actually, for each fixed  $\vxi$,   equations \eqref{Fprop1} are two ordinary parametric equations, in which however the gain profile not only depends on $\vxi$, but also drifts along $x$ and $t$  while  propagating, reflecting  the spatial and temporal  walk-off of the pump wave.  Although it is not difficult to  find complete solutions of  Eqs. \eqref{Fprop1}, this will be done elsewhere \cite{Gatti2024}. With the aim of deriving  manageable results, in this work %we neglect the effects of walk-off and GVM by making  
we make the  more restrictive assumption: 
\beq
\apc (\vxi + \Dxisp\frac{z}{l_c}) \simeq \apc (\vxi) \label{nowoff}
\eeq
The  condition \eqref{nowoff} should not be taken too strictly: as shown in  \cite{Gatti2024}, as long as the pump duration and cross section  are not smaller  than   $\taugvm$  and $\lwoff$,  the effects of the spatio-temporal walk-off do not drastically change the picture, and can be accounted for by an effective gain parameter. 
\par
Assuming the validity of \eqref{nowoff}, the solution of Eqs.\eqref{Fprop1}, with initial conditions $\f_1 (\w,\vxi)|_{z=0}=1$ and  $\f_2 (\w,\vxi)|_{z=0}=0$,  can be readily found. At the crystal exit face $z=l_c$ it reads: 
\beq
\left.\f_j (\w,\vxi\,)\right|_{z=l_c}= e^{-i D(\w) \frac{ l_c}{2}} \F_j (\w,\vxi \,)  \;\;   j=1,2\, ,   \quad\quad \text{with}
\eeq
\beq
\begin{aligned}
\F_1 (\w,\vxi) &=\cosh{\Gamma ( \w,\vxi \,) }+ i \frac{D (\w)l_c}{2 \Gamma(\w,\vxi \,)} \sinh{\Gamma (\w,\vxi \,) }, \\
\F_2 (\w,\vxi) &= g \apc (\vxi)\, \frac{ \sinh{\Gamma (\w,\vxi \,)}}{\Gamma(\w,\vxi\,)}, \\
\Gamma (\w,\vxi \,)&=\sqrt{| g \apc (\vxi)|^2 -\frac{[ D (\w)l_c]^2}{4} }. 
\label{Fresult}
\end{aligned}
\eeq
Finally, the  Bogoljubov transformation \eqref{Bogu} can be recast in terms of these functions using Eq. \eqref{fdef} and \eqref{ajdef}. At $z=l_c$, it becomes the input-output relation:
\beq
\Aout (\w) = e^{i \phi (\w)} \int \frac{d^3\vxi}{(2\pi)^{\frac{3}{2}}} e^{-i \w\cdot \vxi} \left[ \F_1 (\w, \vxi)  \Ain (\vxi) +  \F_2 (\w, \vxi)  \Aindag (\vxi)  \right]
\label{inout}
\eeq
where $\Aout(\w)=\As (\w,l_c) $, $\Ain$ are vacuum input fields and 
\beq
\phi (\w)= \frac{l_c}{2} \left[ k_p +k_{sz} (\w) -k_{sz} (-\w) \right] \label{phi}
\eeq
Some remarks are in order: 
\\
$\bullet$ For the functions $\F_j (\w,\vxi) $, our Ansatz 1 translates into the requirement that they vary on the {\it slow scale} as functions of $\w$, while their Fourier transform with respect to $\vxi$ must decay on the {\it fast scale} (i.e. they are "fairly flat" functions of $\vxi$), so that 
  $ \int d^3\vxi \F_j (\w,\vxi) e^{-i \w_0 \cdot \vxi} \simeq  \int d^3\vxi \F_j (\w \pm \w_0,\vxi) e^{-i \w_0 \cdot \vxi} $. Since these functions depend on $\vxi$ only through the pump profile, and depend on $\w$ only through $\D(\w)l_c $, as anticipated, it is necessary that the pump spectrum is narrow enough with respect to the phase matching bandwidths \eqref{PMbandwidth}, but how much is enough, it depends on the parameter $g$ (see Supplementary S.II for a discussion).\\
$\bullet$ As can be easily verified, 
 $| \F_1 (\w, \vxi) |^2 -
  |\F_2 (\w, \vxi) |^2 =1$. This ensures the unitarity of the tranformation   \eqref{inout}, but only within the limits imposed by our initial Ansatz, as discussed in Supplementary S.I. \\
 $\bullet$ These results are essentially equivalent to those in the Appendix A of \cite{Brambilla2004},  but i) we did not make the narrow frequency band approximation used in \cite{Brambilla2004} ii) consistently, the formalism is different as it is adapted to describe broadband PDC; iii) the approximations made have been clarified.
 %%%%%%%%%%%%%%%%%%
 \subsection{Correlation and coherence functions}
 The  input-output relation \eqref{inout} can be used to calculate
 the two second order moments that for such a Gaussian process  determine all the statistical properties of the PDC emission:
 \begin{subequations}
\begin{align}
 \Psi (\w, \wpr )  = 
 \left\langle \Aout (\w) \Aout ( \wpr)  \right\rangle 
= \,&  e^{i [\phi (\w) +\phi(\wpr)]   }
\int   \frac{d^3\vxi}{(2\pi)^{3}}   e^{-i (\w+\wpr)\cdot \vxi} F_1(\w,\vxi) F_2(\wpr, \vxi)\nn  \\
 \simeq \, & e^{i [k_p  +(\Omega+\Omega^\prime)  k'_s ] l_c}
\int   \frac{d^3\vxi}{(2\pi)^{3}}   e^{-i (\w+\wpr)\cdot \vxi} F_1(\w,\vxi) F_2(-\w, \vxi) 
\label{Psi}  \\
 \Gone (\w, \wpr) =
\left\langle \Aoutdag (\w) \Aout (\wpr)  \right\rangle  
 &=   e^{i [\phi (\wpr) -\phi(\w)] )  } 
\int   \frac{d^3\vxi}{(2\pi)^{3}}   e^{-i (\wpr-\w)\cdot \vxi} F_2(\w,\vxi) F_2^*(\wpr, \vxi) \nn \\
&\simeq  e^{i  (\Omega'-\Omega)  k'_s l_c}
\int   \frac{d^3\vxi}{(2\pi)^{3}}   e^{-i (\wpr-\w)\cdot \vxi}|F_2(\w, \vxi) |^2 ,
%=e^{i [ k_{sz}(\w_2)-k_{sz} (\w_1) ]l_c } \left\langle \aoutdag (\w_1) \aout \w_2)  \right\rangle  
\label{G1}
\end{align}
\label{Corr}
\end{subequations}
where $\F_1$ and $\F_2$ are given by Eq.\eqref{Fresult}. 
As expected,  the {biphoton amplitude} $ \Psi (\w, \wpr )  %= \langle \Aout (\w) \Aout ( \wpr)  \rangle 
$, defining the probability amplitude of finding an entangled photon pair in modes $\w$ and $\wpr$  is peaked at $\wpr=-\w$. Conversely,  the {coherence function} $\Gone (\w, \wpr) %= \langle \Aoutdag (\w) \Aout (\wpr)  \rangle  
$ is peaked at $\wpr=\w$.  Coherently with the requirement of Ansatz 1,  the correlation and coherence peaks decay on the  { \it fast scale},  
so that the second lines [Eqs.\eqref{Psi} and \eqref{G1}] approximately hold (see also the discussion after Eq.\eqref{inout}). 
The phases $\phi (\wpr) \pm \phi (\w)$  were also  approximated  retaining only the dominant terms in the fast variable $\wpr \pm \w$, in agreement with our Ansatz 2. One can easily recognize that  that these phase factors simply accounts for the finite time $k_s' l_c$ taken by the light pulses to cross the medium;
%fact that the fluorescence pulse exits the medium at a time $k_s' l_c$ after the pump pulse entered the medium. 
in what follows we get rid of them by redefining $ t=0$ as the time at which the pump pulse {\em exits} the medium (notice that because of the assumption \eqref{nowoff} this model is not able to account for any GVM  delay). 
\par 
The results in Eqs. \eqref{Psi} and \eqref{G1} can be used in this form to calculate all the properties of the fluorescence emission at  any gain. 
However a further very useful simplification can be performed. Focusing for example on the coherence function \eqref{G1}, in order to study  the mutual coherence  as a function of the difference between  arguments $\w_0=\wpr-\w$,  at a given $\w$, it  would be tempting   to consider a normalized ``distribution'' of the form: 
\beq
\mu_{coh} (\w_0,\w)=
\frac{ \Gone (\w, \w +\w_0) }{ \int d^3 \w_0  \,  \Gone (\w, \w +\w_0)} = \int   \frac{d^3\vxi}{(2\pi)^{3}} e^{-i  \w_0\cdot \vxi}
\frac{ |F_2(\w, \vxi) |^2 }{ | F_2 (\w,0)|^2} 
\eeq
As shown in detail in Supplementary S.III,  in the quasi-stationary conditions this quantity basically do not depend on $\w$ for all the modes within the phase-matching bandwidth, and can be replaced by its peak value at  $\D(\w)=0$. Namely, we have: 
\beq
\frac{| F_2 (\w, \vxi)|^2  }{| F_2(\w,0)|^2}  \simeq \frac{\sinh^2{[g \apc(\vxi)]}  }{\sinh^2{g}} : =\Fcoh (\vxi), 
\label{Fcoh}
\eeq
and similarly: 
\beq
\frac{F_1 (\w, \vxi) F_2(-\w,\vxi) }{F_1(\w,0) F_2(-\w,0)}  \simeq \frac{\cosh{[g \apc(\vxi)]} \sinh  {[g \apc(\vxi)]} }{\cosh {g} \sinh{g}} : =\Fcorr (\vxi) 
\label{Fcorr}
\eeq
for all the modes $\w$ within the phase matching bandwidth. 
In this way, the coherence and correlation functions assume the factorized form that was proposed, with heuristic arguments, in Ref.\cite{Gatti2023}: 
\begin{subequations}
\begin{align}
&\psi (\w,\wpr) =  e^{i k_p l_c}  \F_1(\w,0) \F_2(-\w,0) 
\int   \frac{d^3\vxi}{(2\pi)^{3}}   e^{-i (\w+\wpr)\cdot \vxi}  \Fcorr (\vxi)   , 
\label{PsiSR}   \\
&\Gone (\w,\wpr) = \left| \F_2(\w,0)\right|^2
\int   \frac{d^3\vxi}{(2\pi)^{3}}   e^{-i (\wpr-\w)\cdot \vxi}  \Fcoh (\vxi)    , 
\label{G1SR}
\end{align}
\label{CorrSR}
\end{subequations}
\noindent 
where $ \Fcorr $ and $\Fcoh$ are given by Eqs. \eqref{Fcorr} and \eqref{Fcoh}. We notice that  $F_1(\w, 0)$  and $F_2 (\w,0)$ coincide with  the well know functions of the PWP solution\cite{Klyshko1988}, % $ U(\w)$ and $ V(\w)$, 
while the Dirac-delta correlation of the PWP model are now  replaced by finite peaks: 
\beq
\mu_\beta (\w_0) = \int   \frac{d^3\vxi}{(2\pi)^{3}}   e^{-i \w_0\cdot \vxi}  \F_\beta (\vxi)  , \quad (\beta=corr, \, coh)
\label{mubeta}
\eeq
%$\mucorr(\w_0) = \int   \frac{d^3\vxi}{(2\pi)^{3}}   e^{-i \w_0\cdot \vxi}  \Fcorr (\vxi)   $ 
%and  $\mucoh(\w_0) = \int   \frac{d^3\vxi}{(2\pi)^{3}}   e^{-i \w_0 \cdot \vxi}  \Fcoh (\vxi)   $. 
We remark that depending on the application both expressions \eqref{Corr} or   \eqref{CorrSR} can be used: 
clearly, the latter form is  more manageable for analytical calculations, due to its factorized form. 

%%%%%%%%%%%%%%%%%%%%%%%%%%%%%%%%%%%%%%%%%%%%%%%%%%%%%%%%%%%%%%%%%%%%%%%%%%%%%%%%%%%%%%%%
%%%%%%%%%%%%%%%%%%%%%%%%%%%%%%%%%%%%%%%%%%%%%%%%%%%%%%%%%%%%%%%%%%%%%%%%%%%%%%%%%%%%%%%%
\section{Connection with experimental observations}
The quasi-stationary  model derived in the previous section (hereinafter  QS model) is approximate,
  but it has the advantage of describing, through relatively simple formulas, all aspects of parametric generation, both at a classical and quantum level and in any gain regime.
%proposing thus a unified description of the process.
 In the following we present a  (non-exhaustive)   review of   experimental  observations of high-gain PDC performed   in the last 20  years and of their connection with the  QS model. 
\subsection{Spectral properties (angle-frequency domain)}
This section is devoted to the properties of  parametric radiation in the Fourier domain, which in experiments corresponds to the angle-frequency domain. 
\par
\begin{figure}[hb]
\centering
{{\includegraphics[width=0.65\linewidth]{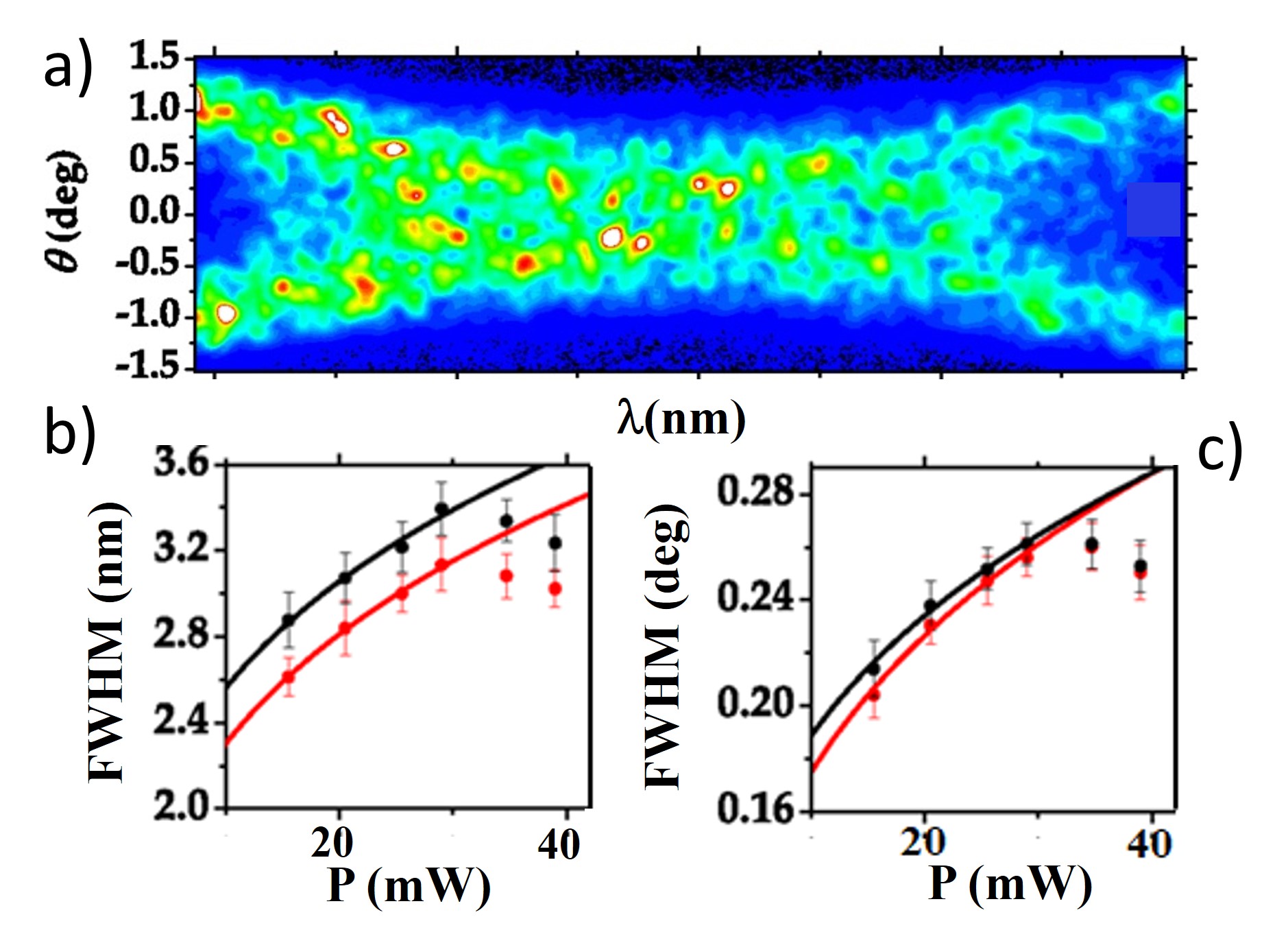}}}
\caption{(a) Example of single shot $(\theta, \lambda)$  spectrum of the twin beams generated in a type I BBO crystal pumped by a 4.5 ps pulse at 349 nm; spectral (b) and angular (b) FWHM of the intensity autocorrelation (red circles) and cross-correlation (black circles)  peaks as function of the pump power. The solid lines are a fit carried out out for the first four  data points under the assumption of undepleted pump beam \cite{Allevi2014}.}
\label{fig_PRA2014}
\end{figure}

Regarding the coherence and correlation peaks $\mu_\beta$ defined in Eq.\eqref{mubeta}, the QS model predicts a substantial broadening of their sizes  for increasing gain. This was extensively discussed in \cite{Gatti2023} and is further detailed in Supplementary  S.II. 
The highlight result  (see Fig.S2) is that the coherence and correlation sizes in each direction of the Fourier space increase with gain proportionally to  $ \sqrt{\frac{g}{\tanh(g)}}$
and $\sqrt{\frac{2g}{\tanh(2g)}}$,  respectively. Notice that, rather  than the second order field moments \eqref{Corr}, experiments typically measure the correlations of light intensity, which are connected to the former by the Gaussian factorization theorem: $\langle \delta\hat I (\w)\delta\hat I(\wpr)\rangle = \langle \hat I (\w)\rangle \delta(\w-\wpr)+ |\Gone(\w, \wpr) |^2 + \left|\Psi (\w, \wpr)\right|^2$, where 
$\delta \hat I (\w) = \Aoutdag (\w) \Aout (\w)-\langle\Aoutdag (\w) \Aout (\w)\rangle$. %, and $:\;:$ indicates normal ordering.   
The first  term represents the shot-noise, while  the two other terms at r.h.s. correspond to the classical  {\it auto-correlation} peak,  centered at $\w=\wpr$ and to   a  {\it cross-correlation} peak,  centered at $\wpr= -\w$, of purely quantum origin.
%Figs.\ref{fig_JMO2006}a,b. display for two different gains the cross-correlation peak of this correlation function for $\vec{q}_2=-\vec{q}_1$, corresponding to symmetrical propagation directions close to degeneracy, a clear manifestation of the transverse momentum entanglement of the emitted twin photons which can be observed in the source far-field (see \cite{Jedr2006} for more details). Notice that the same correlation function also exhibits an auto-correlation peak for $\vec{q}_2=\vec{q}_1$ (not shown in the figure).
\\
With the use of CCD cameras with their high pixel resolution, several experiments  were able to detect the speckled spatial pattern of the radiation generated by high-gain PDC, realized  by pumping a nonlinear crystal with  pulses typically  in the picosecond range. In \cite{Jedr2004} and  \cite{Jedr2006} a first description of the spatial properties of PDC radiation emitted by a  type II  BBO crystal pumped by a Nd:glass laser (1ps, 20 Hz) was given. A huge number of radiation transverse modes was detected  by means of a high quantum efficiency CCD camera placed in the far-field of the source. The single-shot signal and idler images  revealed the speckle-pattern aspect of the far-field radiation, and  directly showed an enlargement of the speckles dimensions (corresponding to the coherence area of PDC) as a function of the pump intensity, i.e. of the gain. Most importantly, this pioneering 
work allowed to experimentally demonstrate -for the first time to our knowledge- a sub-shot noise quantum correlation between a huge number of coupled  transverse modes.
%%%figura JMO2006 %%%%%%%
\begin{figure}[h]
\centering
{{\includegraphics[width=0.95\linewidth]{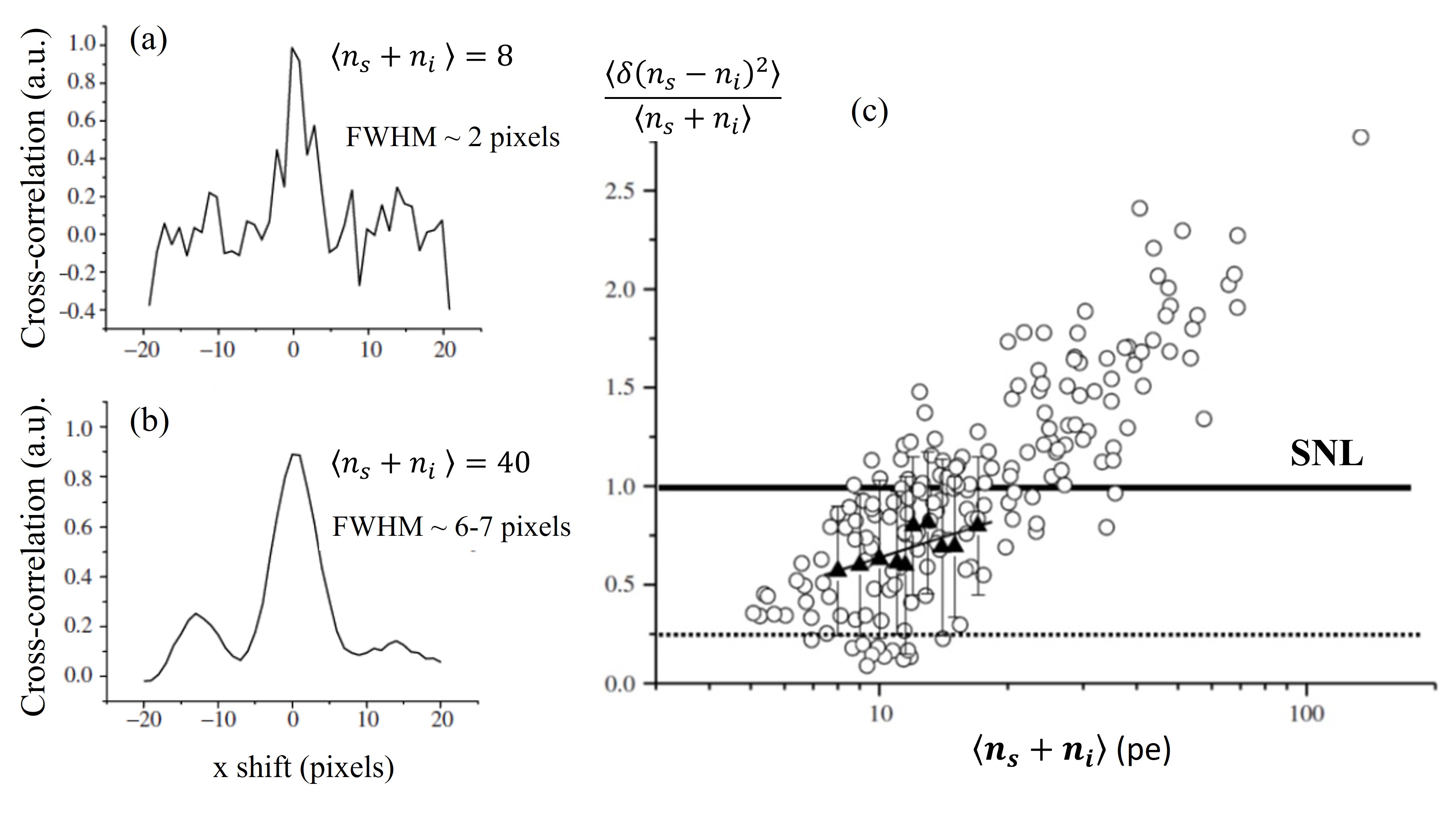}}}
\caption{(a) and (b): Profile of the spatial cross-correlation of photon-numbers measured in  \cite{Jedr2006}, as a function of the  transverse shift from symmetrical far-field positions; (c) The variance of the photon-number difference $  n_s-  n_i $ detected from  symmetric signal-idler pixels, normalized to the shot noise $\langle n_s+n_i \rangle$ is plotted as a function  of $\langle n_s+n_i \rangle$ \cite{Jedr2004}.}
\label{fig_JMO2006}
\end{figure}
%%%%
 As depicted by  Fig.\ref{fig_JMO2006}c,  the twin-beam character of spatial modes was demonstrated by measuring the variance of the photoelectron difference $n_s-n_i$ of the signal-idler symmetrical pixels versus the mean number  $\langle n_s+n_i \rangle$ of the down-converted photons of the pixel-pair. As discussed in \cite{Jedr2006}, the increase of   the transverse coherence area  with gain  is associated with an increase of the transverse correlation area,  as shown in Figs.\ref{fig_JMO2006}a and b. 
 Then, as a consequence of this broadening, a transition from the quantum to the classical regime of the measured pixel by pixel correlation  was observed \cite{Jedr2004}, also confirmed by subsequent experiments performed by Brida et al. \cite{Brida2009}.\\
A systematic study of the transverse spatial coherence  of high gain PDC was performed by Brida et al. in \cite{Brida2009bis}. In that work, a substantial increase of the far-field coherence area with  the mean number of detected PDC photons,  in turn related to the  gain, was observed, thus  demonstrating that in high-gain the speckle size does not depend only on the pump diameter,   as in the low-gain regime, but also  on the pump intensity.
\\
The coherence and correlation properties of twin beams were analysed deeply in \cite{Allevi2014}. This work characterized not only the angular features of twin beams, but also their spectral properties, thanks to the use of an imaging spectrometer placed in the far-field, as done in  \cite{Jedr2006a}.
An example of a single-shot $(\theta, \lambda)$ spectrum of the twin beam radiation generated in a type I PDC process and  recorded by an EMCCD camera, is reported in Fig.\ref{fig_PRA2014}a, showing a speckled X-shaped spectrum.
%%%figura PRA2014 Allevi%%%%%%%
%%%%
 In \cite{Allevi2014} the changes of the angular-spectral radiation pattern with the pump power were investigated by evaluating the photon-number correlation coefficient. Figs.\ref{fig_PRA2014}b and c clearly show the increase of the spectral and spatial FWHM size of the intensity autocorrelation (red circles) and cross-correlation (black circles) with increasing pump power, similarly to what observed in \cite{Allevi2014bis}, where the spatio-spectral coherence properties of bright twin beam as a function of different pump beam parameters were investigated.
\par 
Regarding the angle-frequency spectrum,  several works \cite{Spasibko2012, Sharapova2020,Kulkarni2022}  highlighted that it broadens with increasing gain. This is also predicted by 
our QS model: 
by using  Eqs.\eqref{CorrSR}, it is indeed not difficult to provide an analytical (or semi-numerical)  estimate of  the spectral bandwidths,  which is reported in  Section S.II of Supplementary (Fig. S1).  For instance, for type I collinear PDC  the HWHMs of the  intensity spectrum are approximately given by 
$ \frac{ \Delta \Omega}{\Omgvd} = \frac{ \Delta q}{\qdiff}= \sqrt[4]{ \frac{4 g^2 \log{2}}{  g \coth (g) -1}} % \right)^{1/4},
$ 
in the central part of the spectrum. In the arms of the X-spectrum, instead, one has a faster grow 
$
\frac{ \Delta \Omega}{\Omgvd} = \frac{ \Delta q}{\qdiff} =\frac{\Omgvd}{2  \Omega_\textsc{pm}  }
\sqrt{   \frac{4 g^2 \log{2}}{  g \coth (g) -1}} % \right)^{1/2}
$, where $ \Omega_\textsc{pm} \gg \Omgvd $ is the frequency at which phase-matching occurs. 
 Notice that these formulas are obtained from the factorized form of the coherence function  in Eq.\eqref{CorrSR}, and as such they coincide with the results of the PWP model. If more refined results are needed, the less approximated formulas \eqref{Corr} can be used. \\
 The dependence of the spectral bandwidth on the parametric gain was investigated in \cite{Spasibko2012}, for a  range of $g$  between 3.9 and 6.5 (below which the PDC signal was too low to be measured). In that setup, a BBO crystal was pumped at  354.7 nm by  the third harmonic of a Nd:Yag laser, and the pump pulse had 18 ps duration and up to 0.1\,mJ  energy.
%%%figura Spasibko 2014 %%%%%%%
\begin{figure}[h]
\centering
{{\includegraphics[width=0.95\linewidth]{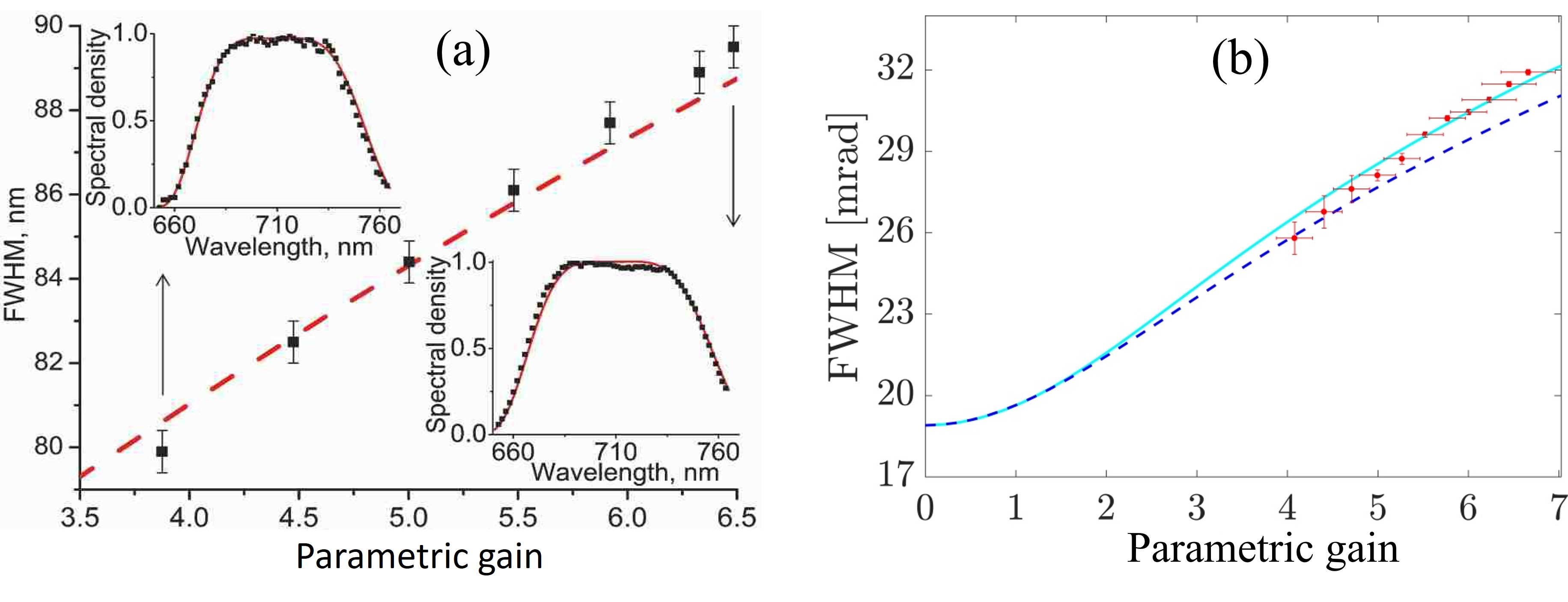}}}
\caption{(a) Dependence of the PDC spectral width on the parametric gain (BBO crystal pumped at 354.7 nm  with a 18 ps pulse). Insets show the shape of the spectrum at gain  3.9 and 6.5 respectively. Reprinted with permission from \cite{Spasibko2012}  \copyright Optical Society of America. (b) FWHM of the angular PDC spectrum reprinted from \cite{Sharapova2020}. Experimental data (red points)  are superimposed to  the theoretical prediction of  the spatial model  (cyan curve) developed in \cite{Sharapova2020} and  of the  the PWP model (dashed curve).}
\label{fig_Spasibko2012}
\end{figure}
%%%%
The behaviour  reported in Fig.\ref{fig_Spasibko2012}a shows the expected broadening of the emitted spectrum of the radiation with the parametric gain.  Clearly, from the reported  data  it is not possible to say whether the grow rate is  the  $\sim \sqrt[4]{g}$  predicted by our QS model, and it would be interesting  to verify with the authors of \cite{Spasibko2012}. 
In the angular domain, a  broadening of the spatial distribution of the far-field  radiation and of the biphoton amplitude was observed and quantified in \cite{Sharapova2020}, where the authors also developed a  theoretical-numerical approach to describe the features of high-gain PDC in the purely spatial domain. Fig.\ref{fig_Spasibko2012}b, taken from \cite{Sharapova2020}, reports e.g. the evolution of the FWHM of the angular intensity distribution of the fluorescence with gain,  measured in a experimental setup similar to that of \cite{Spasibko2012}.
%%%figura Sharapova2020 %%%%%%%
%\begin{figure}[h]
%\centering
%{{\includegraphics[width=0.5\linewidth]{Fig_Sharapova2020.jpg}}}
%\caption{FWHM of the angular intensity distributions of  PDC radiation generated  by pumping a BBO crystal at 354.7 nm  with a 18 ps pulse, reprinted from \cite{Sharapova2020}. Experimental data (red points)  are superimposed to  the theoretical prediction of  the spatial model  (cyan curve) developed in \cite{Sharapova2020} and  of the  the PWP model (dashed curve).}
%\label{fig_Sharapova2020}
%\end{figure}
%%%%
Recent experiments  showing the broadening of the far-field intensity profile and of the  wavelength spectrum  with increasing gain, are reported in  \cite{Kulkarni2022}. 
\subsection{Space-time properties}
\begin{figure}[h]
\centering
{{\includegraphics[width=0.75\linewidth]{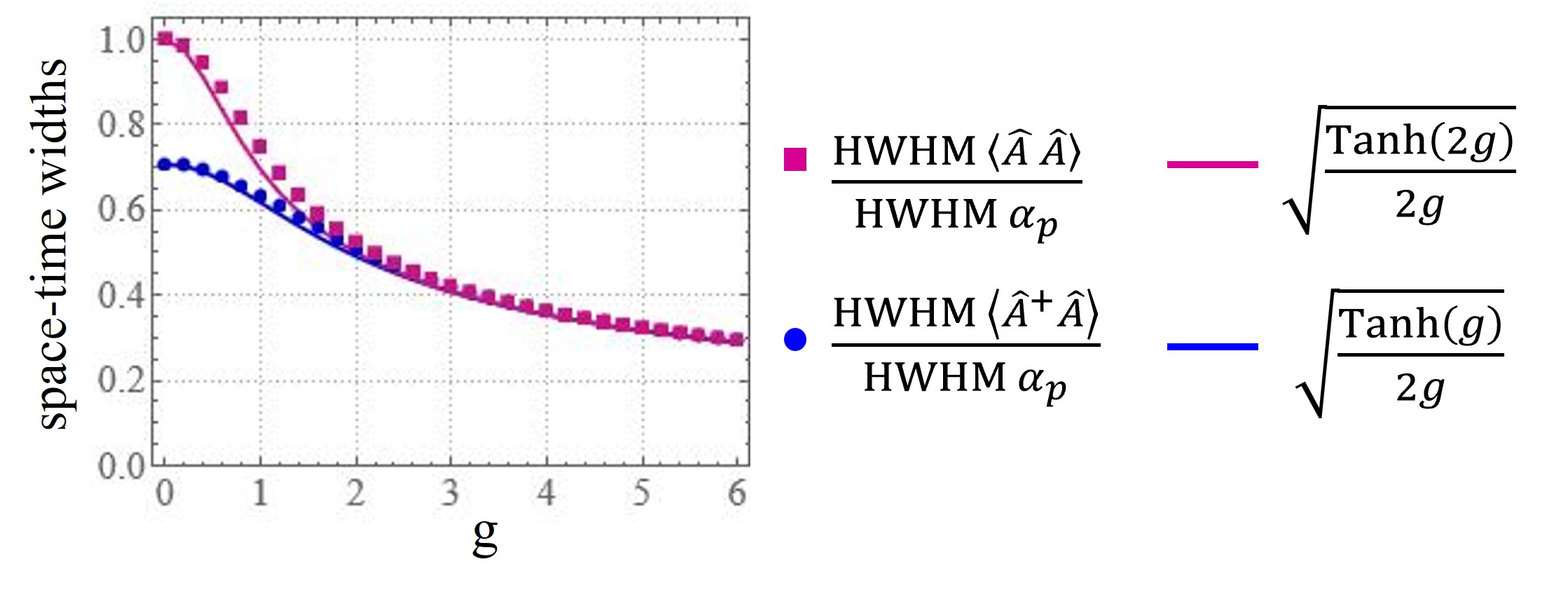}}}
\caption{Theoretical widths HWHM of the photon-number distribution $\Fcoh$,  (blue) and  biphoton amplitude   $\Fcorr $ (purple), in space-time (see  Eqs. \eqref{Fcoh} and \eqref{Fcorr}) as functions  of  $g$. The symbols are numerical evaluations, while the solid lines are analytical estimation from a Taylor expansion of the functions.  }
\label{fig_HWHM}
\end{figure}

This section is devoted to the properties of the fluorescence radiation in the complementary  space-time domain,  typically observed by 
imaging  the crystal output plane \cite{Brambilla2012,Jedr2012a}. 
\\
By Fourier anti-transforming the correlation and coherence \eqref{CorrSR} of the QS model, one obtains:
\begin{subequations}
\begin{align}
&\psi(\vxi, \vxi\,')  = \llangle  \Aout (\vxi) \Aout (\vxi\,') \rrangle = 
 \Fcorr(\vxi)\int\frac{d^3\w}{(2\pi)^3}e^{i\w\cdot(\vxi\,' -\vxi  )}F_1(\w,0)F_2(-\w,0) ,
\label{psi_nf}  \\
&\Gone(\vxi, \vxi\,')  = \llangle  \Aoutdag (\vxi) \Aout (\vxi\,') \rrangle = \Fcoh(\vxi)\int\frac{d^3\w}{(2\pi)^3}e^{ i\w\cdot(\vxi\,' - \vxi)}|F_2(\w,0)|^2  ,
\label{G1_nf}
\end{align}
\label{CorrNR}
\end{subequations}
The Fourier integrals at the r.h.s.   represent space-time correlation and coherence peaks centered at $\vxipr=\vxi$ and  fastly decaying with distance, which will be discussed at the end of this section. 
The slowly varying functions $\Fcorr(\vxi)$ and $\Fcoh(\vxi)$ 
defined by Eqs.\eqref{Fcoh} and \eqref{Fcorr} act as  modulating envelopes. Then,  $\Fcoh $ provides the space-time distribution of the photon number at the crystal output plane, i.e. the transverse and temporal shape of the downconverted pulse. Conversely,  $\Fcorr$ is the biphoton amplitude, with $|\Fcorr|^2$ giving the probability distribution of finding two entangled photons at the space-time point $\vxi$. 
By simple analytical considerations, it is easy to show that  $\Fcorr$ and $\Fcoh$, which at low gain superimpose to the pump amplitude and intensity distribution, at high gain undergo a significant narrowing, with their width  scaling in each direction of space-time proportionally to  
$\sqrt{\frac{\tanh(2g)}{2 g}}$ and $\sqrt{\frac{\tanh(g)}{ g}}$, respectively (see   Fig.\ref{fig_HWHM}). Clearly, this  narrowing of the distributions, 
already partially highlighted in  Ref.  \cite{Gatti2023}, is nothing else than the counterpart of the broadening of the coherence and correlation peaks in the Fourier domain discussed in the previous section. 
To the best of our knowledge,  the only experimental observations related to these aspects were reported by \cite{Tom2023}. 
%%%figura plot HWHM %%%%%%%
%%%%
%%%figura Jedr2012 %%%%%%%
\begin{figure}[h]
\centering
{{\includegraphics[width=0.85\linewidth]{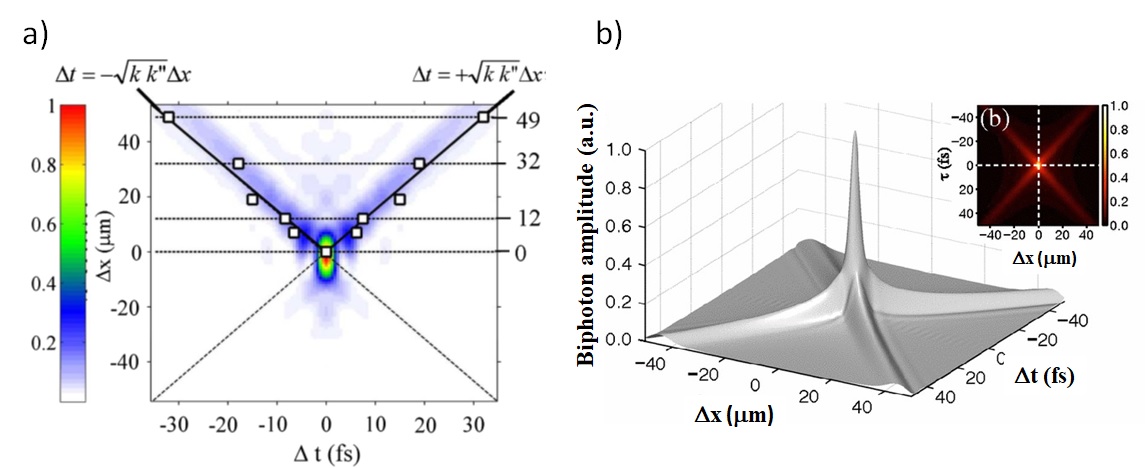}}}
\caption{(a) Biphoton correlation $|\Psi (x,t; x+\Delta x, y+\Delta t)|$  measured by detecting the intensity of  the SFG signal produced  in a second crystal \cite{Brambilla2012,Jedr2012b} as  a function of  a delay $\Delta t $ and of a  transverse shift $\Delta x$ imposed on the twin beams.  Due to the particular setup, only half of the X-shaped correlation was reproduced in the experiment. The experimental data points (open squares) are superimposed  to a theoretical simulation of the experiment (density plot). Reprinted from \cite{Jedr2012b}. The biphoton amplitude calculated \cite{Gatti2009} at the crystal output is shown in (b) for comparison.}
\label{fig_Xexp}. 
\end{figure}
%%%%
\par 
Regarding the space-time correlation and coherence, the Fourier integrals at the r.h.s. of Eqs.\eqref{CorrNR} are essentially the same results as the PWP model. These were  analyzed in \cite{Jedr2007}, for the classical coherence, and in \cite{Gatti2009,Caspani2010} in the case of the quantum correlation. In these works, the characteristic X shape assumed in the space-time domain by the coherence and the correlation functions in proper conditions was highlighted, and the  names “X-coherence” and “X-entanglement” were coined to underline the analogy with the X-waves of classical optics.
\\
The skewed nature in the space-time domain of the coherence  was indirectly measured in \cite{Jedr2007}, where the far-field X-spectrum  in  Fig.\ref{fig_scheme}b), characterized by chromatic and angular dispersion, was measured 
thanks to a spectral diagnostic technique  based on an imaging spectrometer. These results, combined with the generalized form of the well-known  Wiener-Kintchine theorem, clearly indicated that the physical origin of the X coherence  relies on the phase-matching conditions. 
The X-entanglement was instead directly experimentally detected \cite{Jedr2012a,Jedr2012b},  thanks to an interferometric-type scheme based on the inverse process of sum frequency generation (SFG)\cite{Brambilla2012}.  These works on the one side demonstrated the unusually narrow (6 fs FWHM) width of the quantum correlation\cite{Jedr2012a}, which is a consequence of the space-time non factorability\cite{Gatti2009} and  reflects the high frequency-time entanglement of twin beams. On the other side, they  demonstrated \cite{Jedr2012b} the emission of entangled twin photons along the space-time skewed correlation trajectories  
depicted in  Fig.\ref{fig_Xexp}. 
%and 
%%
\section{Conclusions}
This  work  provides a  mathematical derivation of a semi-analytical model for pulsed parametric fluorescence already presented  in \cite{Gatti2023} without a derivation. 
The model is approximate, being based on an assumption of {\it quasi-stationarity}, but has the indisputable advantage of providing a simple and substantially analytical description of all the key aspects of the process, both at the classical and quantum levels, and in any gain regime. As such,  we hope that it may represent a useful tool for researchers in the field.
The more rigorous derivation of the QS model here performed allows  to thoroughly inspect its  limit of validity: essentially, for a transform-limited pump, the main requirements are 
that the pump duration be much longer than the temporal broadening $\sqrt{|k_s''| l_c}$ associated with group velocity dispersion (GVD) and at least on the same order as  the group velocity delay $(k'_p-k'_s)l_c$ between the two waves (see also the discussion in Supplementary S.II). Similarly, in the spatial domain it is necessary that the cross section of the pump is substantially larger than the diffraction broadening during propagation, and not smaller than the transverse walk-off between the two pulses. 
Then, in each specific configuration  these conditions have to be checked (for instance for long crystals they are more difficult to meet, while close to the zero GVM or zero GVD points they are more easily satisfied), but they are usually met in common  setups of  high or medium-gain PDC, either  quantum or classical,  because these are also  the conditions that ensure high efficiency of  nonlinear conversion.  % because they are the typical  conditions under which the nonlinear conversion is efficient.   
\\
In the second part, we explore the main predictions of the model and we connect them to different experimental observations performed in the high-gain regime, where an analytical description was available only within the plane-wave pump approximation.
The connection with the experiments is made by reviewing the  properties of the PDC fluorescence measured in the frequency and  angular  (i.e. in the far field of the source) domains, and the space and time  properties typically observed by imaging the nonlinear crystal output plane.\\
This work focuses on type I quasi-degenerate PDC. Along the same lines,  a quasi-stationary  model can be easily formulated for type II or non-degenerate PDC, but we stress that the limits of validity of the model should be properly studied in each specific configuration.
\ack{The authors acknowledge support from the project  PRIN 2022K3KMX7 of the MUR, ELISE  "Enhancing multiphoton Light-matter Interactions with Space-time Entanglement",  CUP B53D23005150006.}
%%%%%%%%
\bibliographystyle{RS}
\bibliography{biblio_2024}

\begin{thebibliography}{99}

\bibitem{Burnham1970}
Burnham DC, Weinberg DL. 1970  Observation of Simultaneity in Parametric
  Production of Optical Photon Pairs. {\em Phys. Rev. Lett.} \textbf{25},
  84--87.
(\href{http://dx.doi.org/10.1103/PhysRevLett.25.84}{10.1103/PhysRevLett.25.84})

\bibitem{Law2000}
Law CK, Walmsley IA, Eberly JH. 2000  Continuous Frequency Entanglement:
  Effective Finite Hilbert Space and Entropy Control. {\em Phys. Rev. Lett.}
  \textbf{84}, 5304--5307.
(\href{http://dx.doi.org/10.1103/PhysRevLett.84.5304}{10.1103/PhysRevLett.84.5304})

\bibitem{Eberly2004}
Law CK, Eberly JH. 2004  Analysis and Interpretation of High Transverse
  Entanglement in Optical Parametric Down Conversion. {\em Phys. Rev. Lett.}
  \textbf{92}, 127903.
(\href{http://dx.doi.org/10.1103/PhysRevLett.92.127903}{10.1103/PhysRevLett.92.127903})

\bibitem{Gatti2012}
Gatti A, Corti T, Brambilla E, Horoshko DB. 2012  Dimensionality of the
  spatiotemporal entanglement of parametric down-conversion photon pairs. {\em
  Phys. Rev. A} \textbf{86}, 053803.
(\href{http://dx.doi.org/10.1103/PhysRevA.86.053803}{10.1103/PhysRevA.86.053803})

\bibitem{Gatti2008}
Gatti A, Brambilla E, Lugiato L. 2008  Chapter 5. Quantum imaging. , Progress
  in Optics, vol.~51,  pp. 251--348. Elsevier.
(\href{http://dx.doi.org/https://doi.org/10.1016/S0079-6638(07)51005-X}{https://doi.org/10.1016/S0079-6638(07)51005-X})

\bibitem{Genovese2016}
Genovese M. 2016  Real applications of quantum imaging. {\em Journal of Optics}
  \textbf{18}, 073002.
(\href{http://dx.doi.org/10.1088/2040-8978/18/7/073002}{10.1088/2040-8978/18/7/073002})

\bibitem{Padgett2019}
Moreau P, Toninelli E, Gregory T, J. PM. 2019  Imaging with quantum states of
  light. {\em Nature Reviews Physics} pp. 367--380.
(\href{http://dx.doi.org/10.1038/s42254-019-0056-0}{10.1038/s42254-019-0056-0})

\bibitem{Tabakaev2021}
Tabakaev D, Montagnese M, Haack G, Bonacina L, Wolf JP, Zbinden H, Thew RT.
  2021  Energy-time-entangled two-photon molecular absorption. {\em Phys. Rev.
  A} \textbf{103}, 033701.
(\href{http://dx.doi.org/10.1103/PhysRevA.103.033701}{10.1103/PhysRevA.103.033701})

\bibitem{Raymer2021b}
Raymer MG, Landes T, Marcus AH. 2021  Entangled two-photon absorption by atoms
  and molecules: A quantum optics tutorial. {\em Journal of Chemical Physics}
  \textbf{155}.
(\href{http://dx.doi.org/10.1063/5.0049338}{10.1063/5.0049338})

\bibitem{Schlawin2018}
Schlawin F, Dorfman KE, Mukamel S. 2018  Entangled Two-Photon Absorption
  Spectroscopy. {\em Acc. Chem. Res.} \textbf{51}, 2207–2214.

\bibitem{Ghosh1986}
Ghosh R, Hong CK, Ou ZY, Mandel L. 1986  Interference of two photons in
  parametric down conversion. {\em Phys. Rev. A} \textbf{34}, 3962--3968.
(\href{http://dx.doi.org/10.1103/PhysRevA.34.3962}{10.1103/PhysRevA.34.3962})

\bibitem{Atature2002}
Atat\"ure M, Di~Giuseppe G, Shaw MD, Sergienko AV, Saleh BEA, Teich MC. 2002
  Multiparameter entanglement in femtosecond parametric down-conversion. {\em
  Phys. Rev. A} \textbf{65}, 023808.
(\href{http://dx.doi.org/10.1103/PhysRevA.65.023808}{10.1103/PhysRevA.65.023808})

\bibitem{Gatti2009}
Gatti A, Brambilla E, Caspani L, Jedrkiewicz O, Lugiato LA. 2009
  X-Entanglement: The Nonfactorable Spatiotemporal Structure of Biphoton
  Correlation. {\em Phys. Rev. Lett.} \textbf{102}, 223601.
(\href{http://dx.doi.org/10.1103/PhysRevLett.102.223601}{10.1103/PhysRevLett.102.223601})

\bibitem{Sharapova2015}
Sharapova P, P\'erez AM, Tikhonova OV, Chekhova MV. 2015  Schmidt modes in the
  angular spectrum of bright squeezed vacuum. {\em Phys. Rev. A} \textbf{91},
  043816.
(\href{http://dx.doi.org/10.1103/PhysRevA.91.043816}{10.1103/PhysRevA.91.043816})

\bibitem{Klyshko1988}
Klyshko DN. 1988 {\em Photons and Nonlinear Optics}.
New York: Gordon \& Breach Science Pub.

\bibitem{Kolobov1999}
Kolobov MI. 1999  The spatial behavior of nonclassical light. {\em Rev. Mod.
  Phys.} \textbf{71}, 1539--1589.
(\href{http://dx.doi.org/10.1103/RevModPhys.71.1539}{10.1103/RevModPhys.71.1539})

\bibitem{Brambilla2001}
{Brambilla, E.}, {Gatti, A.}, {Lugiato, L. A.}, {Kolobov, M. I.}. 2001  Quantum
  structures in traveling-wave spontaneous parametric down-conversion. {\em
  Eur. Phys. J. D} \textbf{15}, 127--135.
(\href{http://dx.doi.org/10.1007/s100530170190}{10.1007/s100530170190})

\bibitem{Brambilla2004}
Brambilla E, Gatti A, Bache M, Lugiato L. {2004}  {Simultaneous near-field and
  far-field spatial quantum correlations in the high-gain regime of parametric
  down-conversion}. {\em {Phys. Rev. A}} \textbf{{69}}.
(\href{http://dx.doi.org/{10.1103/PhysRevA.69.023802}}{{10.1103/PhysRevA.69.023802}})

\bibitem{Gatti2003}
Gatti A, Zambrini R, San~Miguel M, Lugiato LA. 2003  Multiphoton multimode
  polarization entanglement in parametric down-conversion. {\em Phys. Rev. A}
  \textbf{68}, 053807.
(\href{http://dx.doi.org/10.1103/PhysRevA.68.053807}{10.1103/PhysRevA.68.053807})

\bibitem{Christ2013}
Christ A, Brecht B, Mauerer W, Silberhorn C. 2013  Theory of quantum frequency
  conversion and type-II parametric down-conversion in the high-gain regime.
  {\em New Journal of Physics} \textbf{15}, 053038.
(\href{http://dx.doi.org/10.1088/1367-2630/15/5/053038}{10.1088/1367-2630/15/5/053038})

\bibitem{Kulkarni2022}
Kulkarni G, Rioux J, Braverman B, Chekhova MV, Boyd RW. 2022  Classical model
  of spontaneous parametric down-conversion. {\em Phys. Rev. Res.} \textbf{4},
  033098.
(\href{http://dx.doi.org/10.1103/PhysRevResearch.4.033098}{10.1103/PhysRevResearch.4.033098})

\bibitem{Sharapova2020}
Sharapova PR, Frascella G, Riabinin M, P\'erez AM, Tikhonova OV, Lemieux S,
  Boyd RW, Leuchs G, Chekhova MV. 2020  Properties of bright squeezed vacuum at
  increasing brightness. {\em Phys. Rev. Res.} \textbf{2}, 013371.
(\href{http://dx.doi.org/10.1103/PhysRevResearch.2.013371}{10.1103/PhysRevResearch.2.013371})

\bibitem{Wasilewski2006}
Wasilewski W, Lvovsky AI, Banaszek K, Radzewicz C. 2006  Pulsed squeezed light:
  Simultaneous squeezing of multiple modes. {\em Phys. Rev. A} \textbf{73},
  063819.
(\href{http://dx.doi.org/10.1103/PhysRevA.73.063819}{10.1103/PhysRevA.73.063819})

\bibitem{Gatti2023}
Gatti A, Jedrkiewicz O, Brambilla E. 2023  Modeling the space-time correlation
  of pulsed twin beams. {\em Sci. Rep.} \textbf{13}, 16786.
(\href{http://dx.doi.org/10.1038/s41598-023-42588-y}{10.1038/s41598-023-42588-y})

\bibitem{Brambilla2012}
Brambilla E, Jedrkiewicz O, Lugiato LA, Gatti A. 2012  Disclosing the
  spatiotemporal structure of parametric down-conversion entanglement through
  frequency up-conversion. {\em Phys. Rev. A} \textbf{85}, 063834.
(\href{http://dx.doi.org/10.1103/PhysRevA.85.063834}{10.1103/PhysRevA.85.063834})

\bibitem{LPBbook2015}
Lugiato L, Prati F, Brambilla M. 2015 {\em Nonlinear Optical Systems}.
Cambridge University Press.
(\href{http://dx.doi.org/10.1017/CBO9781107477254}{10.1017/CBO9781107477254})

\bibitem{Jedr2007}
Jedrkiewicz O, Clerici M, Picozzi A, Faccio D, Di~Trapani P. 2007  $X$-shaped
  space-time coherence in optical parametric generation. {\em Phys. Rev. A}
  \textbf{76}, 033823.
(\href{http://dx.doi.org/10.1103/PhysRevA.76.033823}{10.1103/PhysRevA.76.033823})

\bibitem{Jedr2006}
Jedrkiewicz O, Brambilla E, Bache M, Gatti A, Lugiato LA, Trapani PD. 2006
  Quantum spatial correlations in high-gain parametric down-conversion measured
  by means of a CCD camera. {\em Journal of Modern Optics} \textbf{53},
  575--595.
(\href{http://dx.doi.org/10.1080/09500340500217670}{10.1080/09500340500217670})

\bibitem{Allevi2014}
Allevi A, Jedrkiewicz O, Brambilla E, Gatti A, Pe\ifmmode~\check{r}\else
  \v{r}\fi{}ina J, Haderka O, Bondani M. 2014  Coherence properties of
  high-gain twin beams. {\em Phys. Rev. A} \textbf{90}, 063812.
(\href{http://dx.doi.org/10.1103/PhysRevA.90.063812}{10.1103/PhysRevA.90.063812})

\bibitem{Spasibko2012}
Spasibko KY, Iskhakov TS, Chekhova MV. 2012  Spectral properties of high-gain
  parametric down-conversion. {\em Opt. Express} \textbf{20}, 7507--7515.
(\href{http://dx.doi.org/10.1364/OE.20.007507}{10.1364/OE.20.007507})

\bibitem{Jedr2004}
Jedrkiewicz O, Jiang YK, Brambilla E, Gatti A, Bache M, Lugiato LA, Di~Trapani
  P. 2004  Detection of Sub-Shot-Noise Spatial Correlation in High-Gain
  Parametric Down Conversion. {\em Phys. Rev. Lett.} \textbf{93}, 243601.
(\href{http://dx.doi.org/10.1103/PhysRevLett.93.243601}{10.1103/PhysRevLett.93.243601})

\bibitem{Brida2009}
Brida G, Caspani L, Gatti A, Genovese M, Meda A, Berchera IR. 2009a
  Measurement of Sub-Shot-Noise Spatial Correlations without Background
  Subtraction. {\em Phys. Rev. Lett.} \textbf{102}, 213602.
(\href{http://dx.doi.org/10.1103/PhysRevLett.102.213602}{10.1103/PhysRevLett.102.213602})

\bibitem{Brida2009bis}
Brida G, Meda A, Genovese M, Predazzi E, Ruo-Berchera I. 2009b  Systematic
  study of the PDC speckle structure for quantum imaging applications. {\em
  Journal of Modern Optics} \textbf{56}, 201--208.
(\href{http://dx.doi.org/10.1080/09500340802464665}{10.1080/09500340802464665})

\bibitem{Jedr2012a}
Jedrkiewicz O, Blanchet JL, Brambilla E, Di~Trapani P, Gatti A. 2012  Detection
  of the Ultranarrow Temporal Correlation of Twin Beams via Sum-Frequency
  Generation. {\em Phys. Rev. Lett.} \textbf{108}, 253904.
(\href{http://dx.doi.org/10.1103/PhysRevLett.108.253904}{10.1103/PhysRevLett.108.253904})

\bibitem{Gatti2024}
Gatti A. preprint  Effects of spatio-temporal walk-off on pulsed parametric
  generation. alessandra.gatti@ifn.cnr.it.

\bibitem{Jedr2006a}
Jedrkiewicz O, Picozzi A, Clerici M, Faccio D, Di~Trapani P. 2006  Emergence of
  X-Shaped Spatiotemporal Coherence in Optical Waves. {\em Phys. Rev. Lett.}
  \textbf{97}, 243903.
(\href{http://dx.doi.org/10.1103/PhysRevLett.97.243903}{10.1103/PhysRevLett.97.243903})

\bibitem{Allevi2014bis}
Allevi A, Lamperti M, Jedrkiewicz O, Galinis J, Machulka R, Haderka O,
  Pe\v{r}ina J, Bondani M. 2014  Spatio-spectral characterization of twin-beam
  states of light for quantum state engineering. {\em International Journal of
  Quantum Information} \textbf{12}, 1560027.
(\href{http://dx.doi.org/10.1142/S0219749915600278}{10.1142/S0219749915600278})

\bibitem{Tom2023}
Dickinson T, Caspani L, Clerici M et~al.. 2023  Entangled two-photons
  interactions. Public seminar at Insubria University.

\bibitem{Jedr2012b}
Jedrkiewicz O, Gatti A, Brambilla E, Di~Trapani P. 2012  Experimental
  Observation of a Skewed X-type Spatiotemporal Correlation of Ultrabroadband
  Twin Beams. {\em Phys. Rev. Lett.} \textbf{109}, 243901.
(\href{http://dx.doi.org/10.1103/PhysRevLett.109.243901}{10.1103/PhysRevLett.109.243901})

\bibitem{Caspani2010}
Caspani L, Brambilla E, Gatti A. 2010  Tailoring the spatiotemporal structure
  of biphoton entanglement in type-I parametric down-conversion. {\em Phys. Rev
  A} \textbf{81}.
(\href{http://dx.doi.org/10.1103/PhysRevA.81.033808}{10.1103/PhysRevA.81.033808})

\bibitem{handbook}
Gurzadian G, Dmitriev V, Nikogosian D. 1997 {\em Handbook of Nonlinear Optical
  Crystals}.
Springer series in optical sciences. Springer.

\bibitem{Perez2015}
Pérez AM, Spasibko KY, Sharapova PR, Tikhonova OV, Leuchs G, Chekhova MV. 2015
   Giant narrowband twin-beam generation along the pump-energy propagation
  direction. {\em Nature Communications} \textbf{6}.
(\href{http://dx.doi.org/10.1038/ncomms8707}{10.1038/ncomms8707})

\end{thebibliography}
%%%%%%%%%%%%%%%%%
\newpage
\appendix 
 \setcounter{equation}{0}
  \setcounter{figure}{0}
   \setcounter{section}{0}
    \setcounter{subsection}{0}
    \setcounter{page}{1}
  \renewcommand\thepage{S\arabic{page}} 
 \renewcommand\theequation{S\arabic{equation}} 
 \renewcommand\thesection{S.\Roman{section}} 
 \renewcommand\thesubsection{S.\Roman{subsection}} 
  \renewcommand\thefigure{S\arabic{figure}} 
\section*{Appendix: Supplementary Material}
This supplementay information  investigates the range of validity of the approximations made in the main manuscript in order to derive the quasi-stationary model (hereinafter ``QS model'') for multimode pulsed parametric down-conversion (PDC), and provides some specific examples, calculated in realistic PDC setups. 
\\
Section S.I investigates  to what extent  the {quasi-stationary} solution \eqref{inout} of the  propagation equation\eqref{prop} defines a unitary transformation. Consistently, it shows that  unitarity is satisfied only within the assumptions used to derive the model. \\
Section S.II studies the limit of validity of the  Ansatz 1, which requires that the correlation and coherence decay in Fourier space on a much faster scale than the spectral bandwidths, using the results of the {quasi-stationary} model in Eqs.\eqref{CorrSR}. \\
In section S.III we   inspect the limit of validity of  Ansatz 2  of the QS model [ Eq.\eqref{Ansatz2}],   in which the generalized phase matching function $\DD ( \w;   \w_0 - \w)  $ reduces to  $\DD (\w,-\w) - (k'_p -k'_s) \Omega_0 -\frac{\partial k_p}{\partial q_x} q_{0x} $ ). 
\\
The final   section S.IV   studies the validity of the approximations\eqref{Fcoh} and \eqref{Fcorr}, by which the { \em quasi-stationary}   biphoton correlation and the coherence function  in Eq. \eqref{Corr} reduce to their factorized forms of Eqs.\eqref{CorrSR}. 
\subsection{Unitarity of  the {quasi-stationary} solution of the propagation equation} 
In Sec.1(b) we  found a  solution of  the linear  propagation equation \eqref{prop} by using a {\em quasi stationary approximation}, and wrote it as an-input output relation 
\beq
\Aout (\w) = e^{i \phi (\w)} \int \frac{d^3\vxi}{(2\pi)^{\frac{3}{2}}} e^{-i \w\cdot \vxi} \left[ \F_1 (\w, \vxi)  \Ain (\vxi) +  \F_2 (\w, \vxi)  \Aindag (\vxi)  \right]
\label{inoutS}
\eeq
where  $\Ain$ are the  input fields, and  $\F_1$,  $\F_2$ and $\phi$ are defined in Eqs \eqref{Fresult} and \eqref{phi}.  The 
fundamental commutation relation for the output field reads: 
\begin{align}
\left[\Aout (\w), \Aoutdag(\wpr)\right] =& e^{i [\phi(\w)-\phi(\wpr)]} 
\int  \frac{d^3\vxi}{(2\pi)^{3}}  e^{i (\wpr-\w)\cdot \vxi}
\left[ \F_1 (\w, \vxi) \F_1^* (\wpr, \vxi)  - \F_2 (\w, \vxi) \F_2^* (\wpr, \vxi) \right] \label{uni1} \\
  \xrightarrow{\text{Ansatz 1}}&\, 
   e^{i [\phi(\w)-\phi(\wpr)]}   \int   \frac{d^3\vxi}{(2\pi)^{3}}   e^{i (\wpr-\w)\cdot \vxi} \left[ | \F_1 (\w, \vxi) |^2 -
  |\F_2 (\w, \vxi) |^2 \right] \label{uni2} \\
  = &  e^{i [\phi(\w)-\phi(\wpr)]}   \int   \frac{d^3\vxi}{(2\pi)^{3}}   e^{i (\wpr-\w)\cdot \vxi} 1 =  \delta(\w-\wpr) \label{uni}
 \end{align}
 where  the second line holds within  the quasi-stationary  approximation. Indeed, by reformulating  Ansatz1   for the $F_j$, taking into account  the definition \eqref{fdef} and that 
 $
 F_j (\w,\vxi) = e^{i\D(\w) \frac{l_c}{2}}f_j (\w,\vxi),  %=  e^{i\D(\w) \frac{l_c}{2}}  \int  d^3 \w_0 e^{i (\w_0) \cdot \vxi} \Uj (\w, \w - \w_0) 
 $
 it implies   that the Fourier transforms of the 
 $\F_j (\w,\vxi)$ with respect to $\vxi$ die out on the {\em fast scale}. Thus, the r.h.s of Eq.\eqref{uni1} vanishes  for  $\wpr-\w $ outside the fast domain $S_0$,  
 which allows us to replace $\F_j (\w, \vxi) \F_j^* (\wpr, \vxi) 
 \simeq \F_j (\w, \vxi) \F_j^* (\w, \vxi)$ under the integral in \eqref{uni1}. Finally, as can be easily checked, $| \F_1 (\w, \vxi) |^2 -
  |\F_2 (\w, \vxi) |^2 =1$, so that the result \eqref{uni} holds. In a similar way: 
  \begin{align}
&\left[\Aout (\w), \Aout(\wpr)\right] = e^{i [\phi(\w)+\phi(\wpr)]} 
\int  \frac{d^3\vxi}{(2\pi)^{3}}  e^{i (\wpr-\w)\cdot \vxi}
\left[ \F_1 (\w, \vxi) \F_2 (\wpr, \vxi)  - \F_2 (\w, \vxi) \F_1 (\wpr, \vxi) \right] \label{unib1} \\
 &\qquad \xrightarrow{\text{Ansatz 1}}\, 
   e^{i [\phi(\w)+\phi(\wpr)]}   \int   \frac{d^3\vxi}{(2\pi)^{3}}   e^{-i (\wpr+\w)\cdot \vxi} \left[  \F_1 (\w, \vxi) \F_2 (-\w, \vxi) -
  \F_2 (-\w, \vxi)  \F_1 (\w, \vxi) \right]
  =  \, 0 \label{unib}
 \end{align}
  where the last results comes from the fact that $\D(\w)$ is by definition an even function of $\w$, so that $\F_j (-\w)= \F_j(\w) $
%%%%%%%%%%%%%%%%%%%%%%%%%%%%%%%%%%%%%%%%%%%%
\subsection{Ansatz 1: slow and fast scales in Fourier domain.} 
This section studies the limit of validity of the  Ansatz 1, which requires that the correlation and coherence decay in Fourier space on a much faster scale than the spectral bandwidths, using the results of the {quasi-stationary} model in Eqs.\eqref{CorrSR}. 
\par
As for the bandwidths, we consider the two spectral distributions: 
\begin{align}
\Gone (\w,\w) &= \llangle \Aoutdag (\w) \Aout (\w)  \rrangle = \left| \F_2 (\w,0) \right|^2  \mucoh (0) \label{SGone}\\
\Psi (\w,-\w)  &= \llangle \Aout (\w) \Aout (-\w)  \rrangle = \F_1 (\w,0) \F_2 (-\w,0)  \mucorr (0) \label{SPsi}
\end{align}
The first one is the usual light  spectrum (expressed in photon number per unit area and time), giving the probability density of finding a photon in the Fourier mode $\w= (\q,\Omega)$, i.e at frequency $\omega_s +\Omega$ and at external angles $\sin {\theta_x} =\frac{ q_x }{(\omega_s +\Omega)/c}$,  $\sin {\theta_y} =\frac{ q_y }{(\omega_s +\Omega)/c}$.  The second function is such that $ |\Psi (\w,-\w) |^2$  provides the  joint probability distribution of finding two entangled photons  in a correlation volume around the conjugate modes $\w$ and $-\w$, i.e at frequencies $\pm \Omega$ and angles 
 $\sin {\theta_x} =\frac{ \pm q_x }{(\omega_s \pm \Omega)/c}$,  $\sin {\theta_y} =\frac{ \pm q_y }{(\omega_s \pm \Omega)/c}$. 
 The two functions depend on the Fourier coordinates only via the phase-mismatch $\Db (\w) =l_c \D (\w)$. By making a Taylor expansion of these functions in  $\Db $, taken as the independent variable,  around their maxima located at $\Db =0$, one can approximate the decay of the spectral distributions around their peak values as Gaussian functions of the phase mismatch $\Db$: 
 \beq 
\frac{ \Gone (\w,\w)}{ \mucoh (0)  \sinh^2(g)} = 1 -   \frac{g-\tanh (g)}{2 g^2 \tanh (g)} \frac{\Db^2}{2 } + o (\Db^2) \approx 
  e^{-  \frac{g-\tanh (g)}{4 g^2 \tanh ( g)}  \Db^2  } 
  \label{GauG}
 \eeq
  \beq 
\frac{| \Psi (\w,\w) |}{ \mucorr (0)  \sinh(g)  \cosh(g)} = 1 -   \frac{g-\tanh (g)}{2 g^2 \tanh (2 g)} \frac{\Db^2}{2 } + o (\Db^2) \approx 
  e^{-  \frac{g-\tanh (g)}{4 g^2 \tanh (2 g)} \Db^2  } 
  \label{GauPsi}
 \eeq
 Finally, one can use  the Taylor expansion of the phase mismatch \eqref{TaylorPM} to evaluate the spectral widths in  $ q $ and $\Omega$. Introducing the scaled variables  $\bar \Omega =\frac{\Omega}{\Omgvd}$ and $\bar q =\frac{q}{\qdiff}$,    the collinear phase-mismatch parameter  $\Dbz = (2 k_s -k_p) l_c $, and defining $\eta=\mathrm{sgn}(k''_s)$, 
 one has 
 \beq
 \Db (\q,\Omega) \simeq  \Dbz -\bar q^2 + \eta \bar \Omega^2  + o (\bar q^2, \bar \Omega^2)=\begin{cases}
  \bar q_\mathrm{pm}^2-  \bar q^2, \quad  &\text{for }   \bar q_\mathrm{pm}= \sqrt{ \eta \bar \Omega^2 +\Dbz} \in  \mathbb{R}   \vspace{0.4em} \\ 
  \bar \Omega^2 -\bar \Omega_\mathrm{pm}^2,   \quad &\text{for } \bar \Omega_\mathrm{pm}= \sqrt{ \eta( \bar q^2- \Dbz} )\in \mathbb{R}
  \end{cases}
\eeq
where either expression can be used, provided it is real. When e.g. $\eta>0$, for {\em collinear phase matching}  $\Dbz = 0$   both expressions are  always defined,
for  {\em noncollinear phase matching} $\Dbz >0 $ one needs $\bar q > \sqrt{\Dbz}$, while for 
for  {\em nondegenerate phase matching} $\Dbz <0 $ one needs $\bar \Omega > \sqrt{-\Dbz}$ . 
Focusing for example on $\Omega$ (the arguments for the transverse bandwidth are perfectly symmetric), at leading order in the distance from the phase-matching point 
 $ \Delta \bar \Omega=\bar \Omega- \bar \Omega_\mathrm{pm}$ one has: 
$\eta \Db (\w) =
 2 \bar \Omega_\mathrm{pm}  \Delta \bar  \Omega + \Delta \bar  \Omega^2 \to \Delta \bar  \Omega^2 $, for $  \Omega_\mathrm{pm}  \ll  \Delta \bar  \Omega$, and 
 $\eta \Db (\w) \to  2 \bar  \Omega_\mathrm{pm}  \Delta \bar   \Omega $, for $  \Omega_\mathrm{pm} \gg  \Delta \bar  \Omega$. Substituting in the  Gaussian expansions \eqref{GauG}, one can readily evaluate the spectral bandwidth, for example the Half Width at Half Maximum (HWHM) of the photon number distribution \eqref{GauG} as: 
 \beq
\left(\frac{\Delta \Omega} {\Omgvd}\right)_\text{spectr} %=\left(\frac{\Delta q} {\qdiff}\right)_\text{spectr} 
= \left\{ \begin{array}{lc}
 \sqrt[4]{    \frac{4 g^2 \tanh(g) \log{2}}{   g  -\tanh (g)} } %\right)^\frac{1}{4} 
  &\text{for } \bar \Omega_\mathrm{pm} \lesssim 1 %or  \bar q_\mathrm{pm} \ll 1 
  \vspace{0.6em}\\
  \frac{1}{2  \bar  \Omega_\textsc{pm}} 
\sqrt[2]{ \frac{4 g^2 \tanh(g) \log{2}}{   g  -\tanh (g)}} %  \right)^\frac{1}{2} 
&\text{for } \bar \Omega_\mathrm{pm}  \gg 1
\end{array} 
\right.
\label{BandG}
 \eeq
 The same Gaussian approximation provides the  HWHM width of the modulus of the biphoton amplitude \eqref{GauPsi} : 
 \beq
\left(\frac{\Delta \Omega} {\Omgvd}\right)_{|\Psi|} %=\left(\frac{\Delta q} {\qdiff}\right)_\text{spectr} 
= \begin{cases}
 \sqrt[4]{   \frac{4 g^2 \tanh(2 g) \log{2}}{   g  -\tanh (g)}  }  &\text{for } \bar \Omega_\mathrm{pm} \lesssim 1 %or  \bar q_\mathrm{pm} \ll 1 
  \vspace{0.5em}\\
  \frac{1}{2   \bar  \Omega_\textsc{pm}} 
\sqrt[2]{  \frac{4 g^2 \tanh(2g) \log{2}}{   g  -\tanh (g)}  }  &\text{for } \bar \Omega_\mathrm{pm}  \gg 1
\end{cases} 
\label{BandPsi}
 \eeq
 The expressions  for the transverse bandwidth  are completely equivalent, provided one substitutes in the formulas $\frac{\Delta \Omega} {\Omgvd} \to \frac{\Delta q} {\qdiff}$ and $  \bar \Omega_\mathrm{pm}  \to  \bar q_\mathrm{pm} $. 
 %%%%%%%%%%%%%%%%%Fig   S1  %%
\begin{figure}[h]
\centering
{{\includegraphics[width=\linewidth]{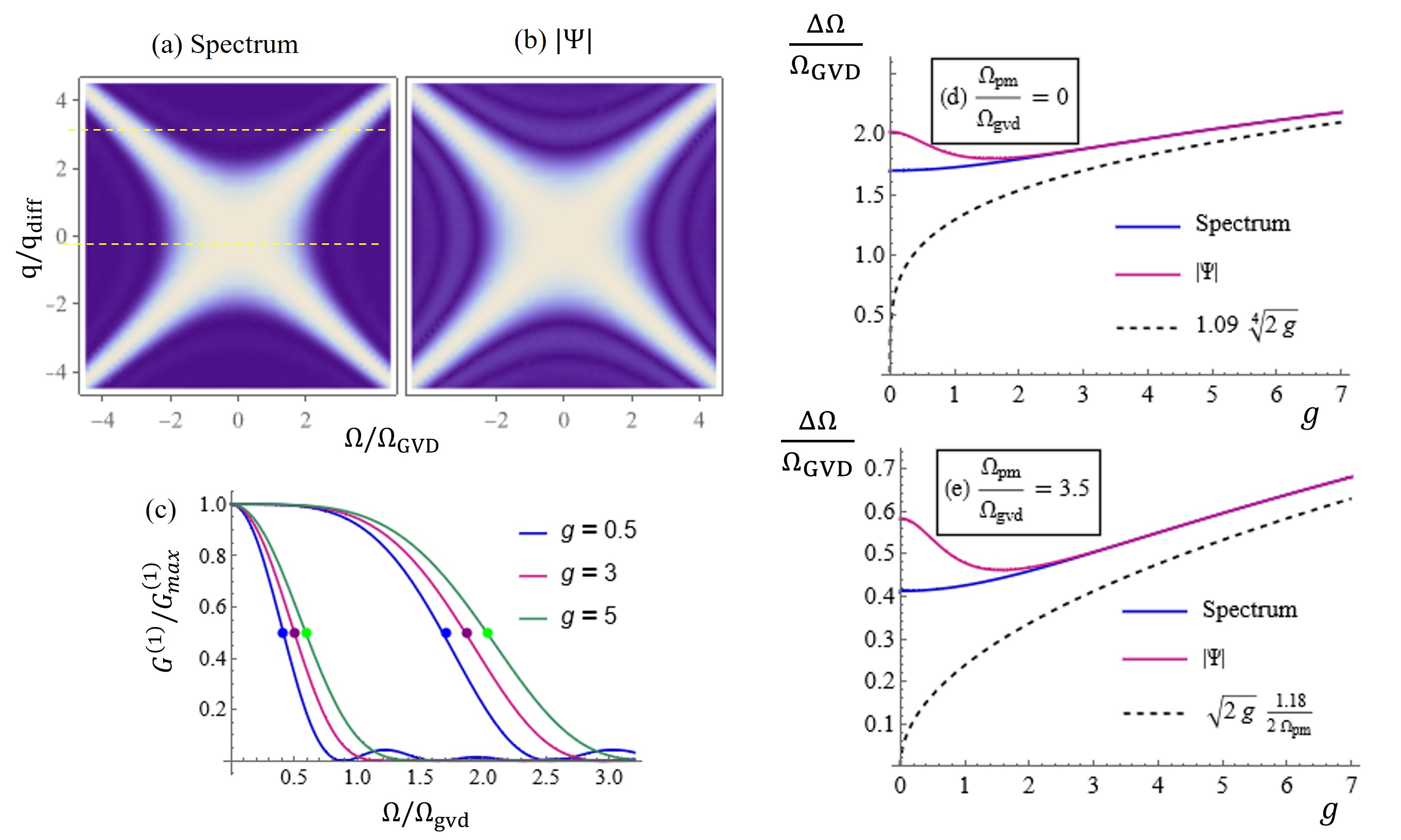}}}
\caption{Spectral bandwidths. (a) and (b): Fourier distributions of  the intensity $\Gone (\w,\w)$ and  of the biphoton amplitude $| \Psi (\w,-\w)|$, from Eqs.\eqref{SGone} and \eqref{SPsi}, respectively, evaluated using the  Sellmeier relation  in \cite{handbook} for a  $2$mm BBO cut for collinear PDC  from $515 \to 1030$nm ($\Dbz=0$), and for $g=1$. (c): Sections of the intensity spectrum along the yellow dashed lines  in (a), for three different values of the gain. The points are the HWHM evaluated from Eqs.\eqref{BandG}. 
(d) and (e) plot  formulas \eqref{BandG} and \eqref{BandPsi} for the HWHM of the distributions, in  (d) the central part of  the spectrum  $\Omega_\text{pm}=0$,  and  e) in   the arms, at $\Omega_\text{pm}=3.5$. The dashed lines are the analytic asymptotes of the curves at large $g$, exhibiting a $\sqrt[4] {2g}$ and $\sqrt{2g}$ grow rate, respectively. 
 The plots for  the transverse bandwidths  are identical,  provided one substitutes  $\frac{\Delta \Omega} {\Omgvd} \to \frac{\Delta q} {\qdiff}$ and $  \bar \Omega_\mathrm{pm}  \to  \bar q_\mathrm{pm} $. 
For these parameters $\Omgvd=107  \text{  THz}$, $\qdiff=71 \text{ mm}^{-1}$. 
%\eqref{GauG} and \eqrf{GauPsi}
}
\label{fig_bands}
\end{figure}
%%%% %%%%%%%%%%%%%%%%%%%%%%%
 These results are plotted by panels (d) and (e) of Fig.\ref{fig_bands}. At low gain, the biphoton amplitude is broader than the intensity spectrum, reflecting the fact that in the spontaneous regime quantum correlated photons originate only from primary down-conversion processes. Therefore, the probability of finding  entangled photon pairs $|\Psi (\w,\w)|^2$  is distributed as the probability of finding single photons $\Gone (\w,\w)$,  so that   $|\Psi (\w,\w)| \propto \sqrt {\Gone (\w,\w)}$ is broader than the intensity spectrum.
  Conversely,  at high gain, also photon pairs that are not generated by the  decay of the same pump photon can be quantum correlated as a consequence  of cascaded stimulated processes, and contribute to the biphoton amplitude.  Hence the two distributions tend to become identical as gain increases. \\
  As  observed by several experiments\cite{Spasibko2012,Sharapova2020}, both Fourier  distributions broaden with increasing gain: our formulas show that the rate of broadening is different in the central part of the spectrum, where asymptotycally the bandwidths grow as $g^{1/4}$, and along the thin arms of the X-shape,  where they grow more rapidly as $g^{1/2}$.  For comparison, the figure also plots the spectral distribution of the intensity  \eqref{SGone} (blue line)  and biphoton amplitude \eqref{SPsi} (purple line), in the concrete example of  a  $2$mm Beta-Barium-Borate (BBO) crystal,  pumped for collinear (i.e. $\Delta_0=0$) down-conversion from  $515 \to 1030$nm, 
  evaluated using the complete Sellmeier relation  in \cite{handbook}. 
  In this example the crystal is cut for collinear and degenerate emission, but we remark that the results of  Fig.\ref{fig_bands}d may also apply to noncollinear phase matching close to the degenerate frequency, while the plot in Fig. \ref{fig_bands}e also describes the widths along $q$ for noncollinear phase matching at $q_\text{pm} = 3.5 \qdiff$, which for our parameters corresponds to an external emission angle $\sim 2.34$ degrees. 
  \\ 
  \par
\begin{figure}[h]
 \centering
{{\includegraphics[width=0.87\linewidth]{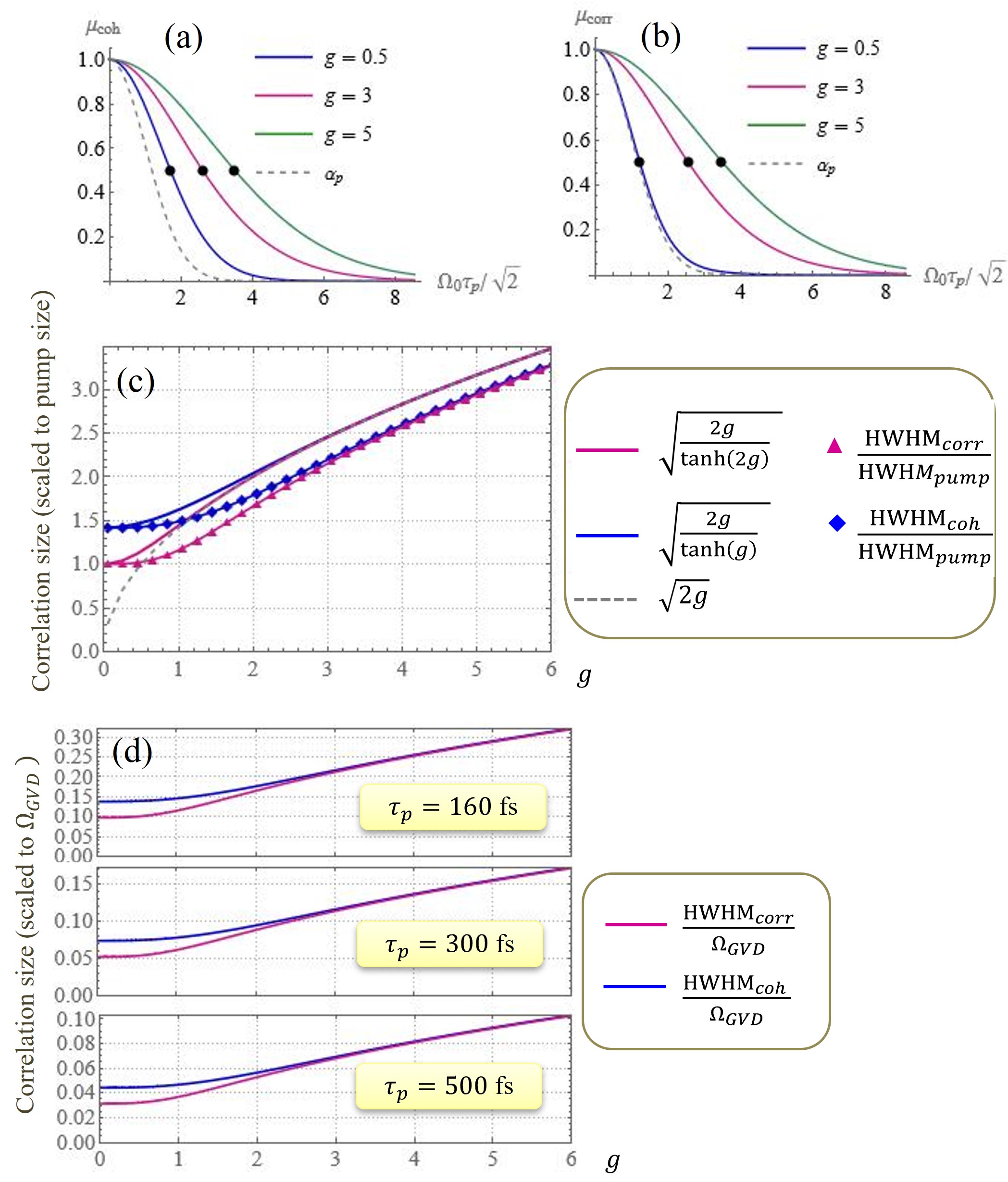}}}
 \caption{Correlation and coherence sizes. (a) and (b): Profile of the coherence and correlation peaks \eqref{mucorr}, as a function of a Fourier coordinate, showing their broadening with increasing gain. (c) Plots the coherence (blue) and correlation (violet) widths, scaled to the pump width,  as a function of the gain. The solid lines are formulas \eqref{sigmacorr}, while the line with symbols are numeric evaluation of the HWHM of the peaks, scaled to the HWHM of $\apc$. (d) Show the correlation widths (HWHM of the peaks), scaled to the characteristic size of the PDC spectrum $\Omgvd$, for three different pump pulse durations. Other parameters as in Fig.\ref{fig_bands}.}
 \label{fig_corr}
 \end{figure}

 
As for the widths of the correlation and coherence in the Fourier domain, we consider the normalized correlation peaks
\beq 
\begin{aligned}
\mucoh (\w_0) &= \int   \frac{d^3\vxi}{(2\pi)^{3}} \frac{ \sinh^2[g\apc(\vxi)]}  {\sinh^2(g)}    e^{-i \w_0\cdot \vxi}  \\
\mucorr (\w_0) &= \int   \frac{d^3\vxi}{(2\pi)^{3}}\frac{ \sinh[2g\apc(\vxi)]}  {\sinh(2g)}     e^{-i \w_0\cdot \vxi}
\label{mucorr}
\end{aligned} 
\eeq
Their widths were evaluated in our Ref.\cite{Gatti2023} as the  mean square deviations 
$  \sigma_{i,\beta}^2 = \int d^3 \w \, w_i^2 \mu_\beta (\w) $, ($i=q_x,q_y,\Omega)$.  %$\beta= corr, coh $, where . 
By assuming a Gaussian pump of the form : 
 \beq 
 \apc ( r,t) = e^{-\frac{r^2}{w_p^2}-\frac{t^2}{\tau_p^2}}
 \eeq
 we obtained: 
 \begin{align}
& \sigma_{i,coh} = \sqrt{\frac{2 g}{\tanh (g)}} \,  \sigma_{i,\text{pump}} , 
&  \sigma_{i,corr} = \sqrt{\frac{2 g}{\tanh (2g)}} \,  \sigma_{i,\text{pump}}
\label{sigmacorr}
 \end{align}
 where $ \sigma_{i,\text{pump}}$ are the widths (root mean square deviation) of the pump Fourier amplitude along the  Fourier coordinates:
 $\sigma_{q,\text{pump}}=\sqrt{2}/w_p\,$, $ \sigma_{\Omega,\text{pump}}=\sqrt{2}/\tau_p$.  Formulas \eqref{sigmacorr} are plotted in  Fig.\ref{fig_corr}c,   along with some examples of the profiles of the coherence peak  (a) and of the the correlation peak (b),  for different gains. 
  %%%  Fig S2 
As  known (see e.g. \cite{Gatti2009,Caspani2010}), at low gain the biphoton correlation profile reproduces the Fourier profile of the pump {\em amplitude}, while the coherence peak is the Fourier transform of the pump {\em intensity}, and hence broader [for a Gaussian pump, $\apc(\Om)\sim \exp{(-\frac{\Omega^2 \tau_p^2}{4})}$, while the Fourier transform of the pump intensity $\sim 
 \exp{(-\frac{\Omega^2 \tau_p^2}{2})}$]. At high gain the two profiles tend to coincide, because of the cross-talk between coherence and correlation arising from stimulated processes. They broaden with the gain, with an asymptotic behaviour $ \sim { g}^\frac{1}{2}$.  The lines with symbols  in figure Fig.\ref{fig_corr}c show a numerical evalution of the HWHM of the peaks, scaled to the HWHM of the pump Fourier profile. Were the peaks Gaussian, these curves would  exactly match the solid lines in the plot. Instead,  they are slightly below, because the correlation peaks have slowly decaying tails.
 \\
  Part (d) of the figure shows instead the correlation width (HWHM of the peaks along the $\Omega_0$ coordinate) scaled to the characteristic spectral bandwidth $\Omgvd$, in the concrete example of collinear PDC in BBO (same parameters as in Fig.\ref{fig_bands})  and for three different pulse durations. 
By comparing with figure \ref{fig_bands} we can conclude that it is not too difficult to satisfy the requirements of the quasi-stationary model (Ansatz1) when one considers  the collinear and degenerate central part of the spectrum, whose width is on the order of $2 \Omgvd$ (we remind that it grows slowly with the gain). In this case the requirement for a Gaussian pump pulse can be summarized as $\tau_p \gg \taugvd=\frac{1}{\Omgvd} = \sqrt{|k''_s l_c|}$. In  our example $\taugvd \simeq 6.6  - 21\, $fs for crystal lengths $1-10\,$mm, and hence it is not difficult to fullfill the requirement for standard pulse durations. Clearly, in the case of a short pulse, the operating conditions must be carefully checked before adopting this model, especially when noncollinear phase matching is used, or when an ultra-broad part of the PDC spectrum,which also includes the thin arms of the "X",  is taken into consideration.
  \par
  The spatial aspects are completely equivalent and can be formulated in terms of $w_p \gg \frac{1}{\qdiff} = \sqrt{\frac{l_c  \lambda}{2 \pi n_s} }\simeq 10-30 \, \mu$m.   We do not discuss in  detail  the spatial aspects because there is normally no reason to use a tightly focused pulse for parametric generation in  the quantum domain, since typically the gain required is not very  high. 
   \par 
 On the other hand, it is important to  notice that the  requirement made in Eq.  \eqref{nowoff},  that   the effects of walk-off and GVM are negligible,  is typically  more restrictive, as it  implies that the pump duration and waist are larger or at  least on the order of $\taugvm$ and $\lwoff$, respectively. Then, the specific operational conditions should be checked: on the one side these quantities scale linearly with the crystal length. On the other,  especially
 $\taugvm$  changes fast with the frequency.  For  example, for a BBO cut for collinear and degenerate phase matching  $k'_p-k'_s$ ranges from $17\,$fs/mm for twin photons generated 
 at $1300\,$nm,   to $355\,$fs/mm for PDC at  $704\,$ nm (it is $92.8\,$fs/mm at $1030$ nm) . The transverse walk-off length $\lwoff $ can be  quite different for different materials and phase matching-conditions, ranging from $\sim 0$  for type 0 or non-critical  phase-matching, to e.g. $\sim 60 \, \mu$m/mm for PDC at $1030$ nm from  a BBO. 
  %%%%%%%%%%%%%%%%%%%%%%%%%%%%%%%%%%%%%%%%%%%%%%%%%%
 \subsection{Ansatz 2: Taylor expansion of phase matching in the fast variables } 
\label{sec:PM}
In this section we inspect the limit of validity of the  Ansatz 2  used to derive the quasi stationary solution,  in which the phase matching function $\DD ( \w;   \w_0 - \w)   =  k_{sz} (\w) +  k_{sz} (\w_0 - \w )  - k_{pz} (\w_0)$ is approximated as 
\beq
\begin{aligned}
\DD ( \w;   \w_0 - \w ) \xrightarrow{\text{Ansatz 2}} \DD_\text{appr}  ( \w;   \w_0 - \w )=  \DD ( \w;   - \w) -\left ( k^\prime_p -  k^\prime_s \right) \Omega_0  -  \frac{\partial k_p}{\partial q_x}  q_{0x}  
\label{Ans2}
\end{aligned}
\eeq
for $\w_0 $ belonging to the {\em fast} domain. \\
To this end, we expand  the phase matching  in Taylor series of the  fast variable  $\w_0 = (q_{0x},q_{0y}, \Omega_0)$: 
%Provided  that   the pump has a frequency-angle spectrum  narrow  enough that its dispersion and diffraction  during propagation are negligibleProvided that the duration and cross section of the pump pulse are large enough  to make  its diffraction and dispersion along the medium negligible  (and/or the crystal is short enough), quadratic and higher order terms can be neglected, so that:  
\begin{align}
l_c [\DD ( \w;   \w_0 - \w) - \DD ( \w;   - \w) ]   & = l_c \left[ \vec{\nabla} k_{sz}  (-\w)-\vec{ \nabla} k_{pz} (0)     \right] \cdot \w_0 + o(q_0
 \lwoff,  \Omega_0\taugvm)  \\
%
%& \simeq \left[ \vec{\nabla} k_{sz}  (-\w)-\vec{ \nabla} k_{pz} (0)    \right] \cdot \w_0 \\
&\simeq l_c  \left (    k^\prime_s  -k^\prime_p -   k''_s \Omega \right) \Omega_0  - l_c  \left( \frac{\partial k_p}{\partial q_x}  \ex  +\frac{\vec q}{k_s}  \right) \cdot \q_0
\label{A3}
\end{align}
where in the second passage only the leading terms of the expansion of $ \vec{\nabla} k_{sz}  (-\w)$ around  $\w=0$  (the central frequency and the collinear direction) have been retained. Here $k^\prime_j $, $ k''_j$ are shorthands for 
 $  \frac{\partial k_j}{\partial \Omega} ({\w=0} )$,  $\frac{d^2 k_j }{d\Omega^2} (\w=0)$, while   %=\frac{1} {v_{gj}} $,   where $v_{gs} $  and $v_{gp} $ denote  the group velocities of the signal and pump wave-packets; 
 $ \frac{\partial k_p}{ \partial q_x} \approx- \rho_p$ gives the   { \em walk-off} angle of the Poynting vector of the extraordinary pump, assumed  here  in the x-direction.  
  The quadratic  terms neglected in Eq.\eqref{A3} are of the form 
  $
  \frac{l_c}{2}( k''_s -k''_p) \Omega_0^2 \simeq \frac{  \Omega_0^2}{2 \Omgvd^2} $   
  and 
    $
 \left(  \frac{l_c}{2 k_s}  - \frac{l_c}{2 k_p}\right) q_0^2 \simeq \frac{  q_0^2}{4 \qdiff^2}$ , where we remind that $\Omgvd=(|k_s''| l_c)^{-1/2} $  and $\qdiff= (k_s/l_c)^{1/2}$ are the characteristic bandwidth of phase matching. Therefore, provided that the request of the quasi-stationary model are satisfied, i.e.  that the correlation functions decay on distances 
 $  \Omega_0 \ll  \Omgvd  $ , $q_0 \ll \qdiff$ these terms can be safely neglected. 
  \\
The term $l_c k''_s \Omega \Omega_0= \frac{\Omega \Omega_0}{\Omgvd^2} $    becomes significant only when the spectral width of the correlation and of the coherence  $\Omega_0$  is larger than 
  $  \approx \Omgvd   \frac{\Omgvd }{\Omega}$.  Clearly, within the request of the quasi-stationary model, this term  is negligible unless the PDC bandwidth considered is very large, so that   $ \frac{\Omega }{\Omgvd}>>1$. \\
For a tightly focused pump, the term $ \frac{\vec q\cdot \q_0}{k_s}  $ may originate the so-called  {\em hot spots} \cite{Perez2015},  which are relevant only close to a   specific angle $q_x/k_s = -\rhop $ (for our BBO $\rho_p=-3.2^\circ$).   In any case, this term is not negligible only 
when $q_0 \approx \qdiff \frac{\qdiff}{q}$, and within the quasi-stationary Ansatz 1, requires that   broad angular bandwidth of PDC  is considered. 
%%%%%%%%%%%%%%%%%%%%
 \begin{figure}[ht]
 \centering
{{\includegraphics[width=\linewidth]{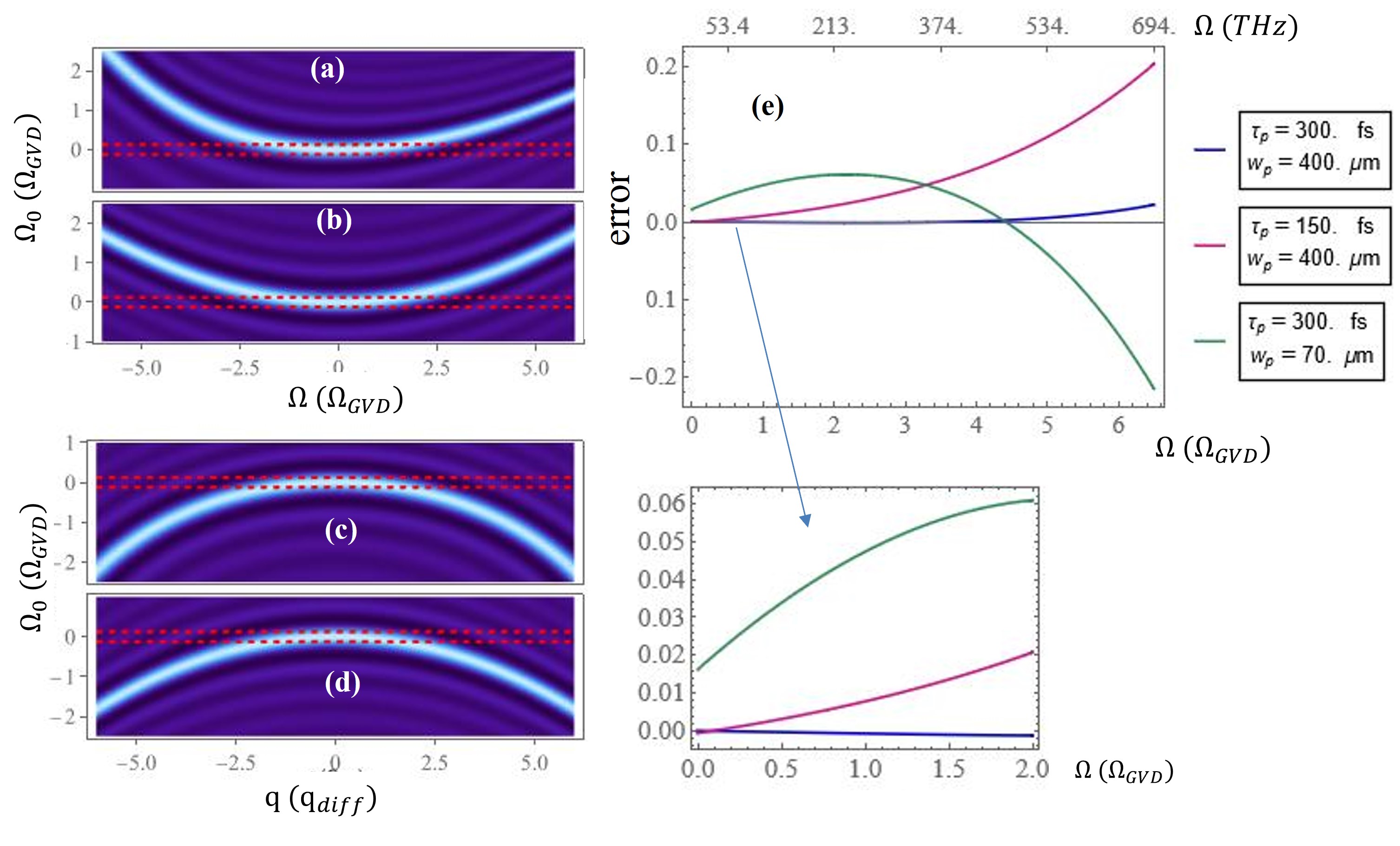}}}
 \caption{(a)  and (b) compare   $\sinc [\DD (\w, \w_0-\w) \frac{l_c}{2} ] $ and $ \sinc[\DD_\text{appr} (\w, \w_0-\w) \frac{l_c}{2}]$, with  $\DD_\text{appr}(\w, \w_0-\w) $  given by  Eq.\eqref{Ans2}, plottted as functions of $\Omega$ (horizontal axis)  and $\Omega_0$ (vertical axis), for $q=q_0=0$. The red dashed lines enclose the region where the Fourier transform of a Gaussian pump of duration $\tau_p=150 fs$ is larger than $1/e$. (c) and (d) are cuts of the same functions at $\Omega=0$ and $q_0=0$.  
 (e) Shows an estimate of the error, plotting  $\sinc[\DD_\text{appr} (\w, \w_0-\w) \frac{l_c}{2}]- \sinc [\DD (\w, \w_0-\w) \frac{l_c}{2}]$, 
  where $\q$ and $\Omega$  are chosen to satisfy   $\D (\q,\Om) =0$, while $\Omega_0=2/\tau_p$ and $q_0=2/w_p$ are the coordinates at which a Gaussian pump $\apc(q_o,\Omega_0) =1/e$. Parameters are those of a 2mm BBO crystal, cut for collinear phase matching at $1030$nm, as in Fig\ref{fig_bands}.}
 \label{fig_comp}
 \end{figure} 
 %%%%%%%%%%%%%%%%%%%%%%%%%%%%%%%%%%%%%%%
\par
  Figure \ref{fig_comp} provides a comparison between the full expression of phase mismatch and its approximate form \eqref{Ans2}, in the example of a BBO cut for collinear and degenerate phase matching at $1030\,$nm. 
  Panels (a) -(d) provide  a  visual  comparison,  showing  density plots of  the functions to be compared. Namely, (a) and (b) plot $\sinc [\DD (\w, \w_0-\w) \frac{l_c}{2} ] $ and $ \sinc[\DD_\text{appr} (\w, \w_0-\w) \frac{l_c}{2}]$, respectively, where  $\DD_\text{appr}(\w, \w_0-\w) $ is given by  Eq.\eqref{Ans2}, as  functions of $\Omega$ (horizontal axis)  and $\Omega_0$ (vertical axis), for $q=q_0=0$. 
  The red dashed lines enclose the region where the Fourier transform of a Gaussian pump of duration $\tau_p=150 fs$ is larger than $1/e$. (c) and (d) plot  2D cuts of  the same functions along $q$ and $\Omega_0$, for a $\Omega=0$ and $ q_0=0$.   From these graphs it is possible to appreciate the validity of the approximation, at least in the central part of the PDC spectrum. In fact, unless the spectral widths of the correlation, which in the first instance we can approximate with those of the pump, are truly of the order or larger than  $\Omgvd$, the approximation is essentially an identity.  Panel (e) shows a more quantitative comparison, plotting 
  \beq
 \sinc[\DD_\text{appr} (\w, \w_0-\w) \frac{l_c}{2}]- \sinc [\DD (\w, \w_0-\w) \frac{l_c}{2} ] 
 \label{error}
  \eeq
  where  the variables $\Omega_0=2/\tau_p$ and $q_0=2/w_p$ are the coordinates at which a Gaussian pump $\apc(q_o,\Omega_0) $ reduces to $1/e$, and $\q$ and $\Omega$ are chosen as to satisfy phase-matching for a plane-wave pump, i.e.  $\D (\q,\Om) =0$. Then,  for $\w_0=0$, we have that $\sinc[\DD_\text{appr} (\w, -\w)\frac{l_c}{2}] =\sinc [\DD (\w, -\w) \frac{l_c}{2} ] =1$, and the quantity in Eq. \eqref{error} is substantially the error normalized to unity. 
  We notice that   even in the case of the red and green curves, for which the pump duration and waist  are rather small (i.e. the corresponding spectral widths are rather large), the error done with the approximation becomes appreciable only when  $\Omega \ge 4  \Omgvd$ in this example, that corresponds to a large PDC bandwith  $\sim 850$ THz or $\sim 500$ nm around $\lambda_s =1030\,$nm.  
  
  %%%%%%%%%%%%%%%%%%%%%%
  \subsection{Factorability of the quasi-stationary correlation and coherence functions.} 
  This section studies the validity of the approximations \eqref{Fcorr} and \eqref{Fcoh}, by which the { \em quasi-stationary}   biphoton correlation and the coherence function  in Eq. \eqref{Corr} reduce to their factorized forms reported by Eqs. \eqref{CorrSR}. 
 \par 
 The approximation is based on the observation  that the kernels $ F_1 (\w,\vxi)$, $ F_2(\w, \vxi)$  
  of the  Bogoljubov relations \eqref{inout}  that solve the propagation equation in the quasi-stationary limit are such that the ratios 
  $$
  \frac{ F_1 (\w, \vxi)  }{ F_1(\w,0)|},   \qquad  \frac{ F_2 (\w, \vxi)  }{ F_2(\w,0)|}
  $$ 
 depend very slowly on the Fourier variable $\w$ within the phase-matching bandwidth, where they  remain close to their peak value attained at $\D (\w)=0$. Therefore, one can  approximately set: 
 \beq
   \begin{aligned} 
& \frac{F_1 (\w, \vxi) F_2(-\w,\vxi) }{F_1(\w,0) F_2(-\w,0)}  \to \frac{\cosh{[g \apc(\vxi)]} \sinh  {[g \apc(\vxi)]} }{\cosh {g} \sinh{g}} : =\Fcorr (\vxi) \, \\
&\frac{| F_2 (\w, \vxi)|^2  }{| F_2(\w,0)|^2}  \to \frac{\sinh^2{[g \apc(\vxi)]}  }{\sinh^2{g}}  =\Fcoh (\vxi)
\label{FcorrS}
\end{aligned}
\eeq
%%%%%%%%%%%%%%%%%%%%
 \begin{figure}[ht]
 \centering
{{\includegraphics[width=0.92\linewidth]{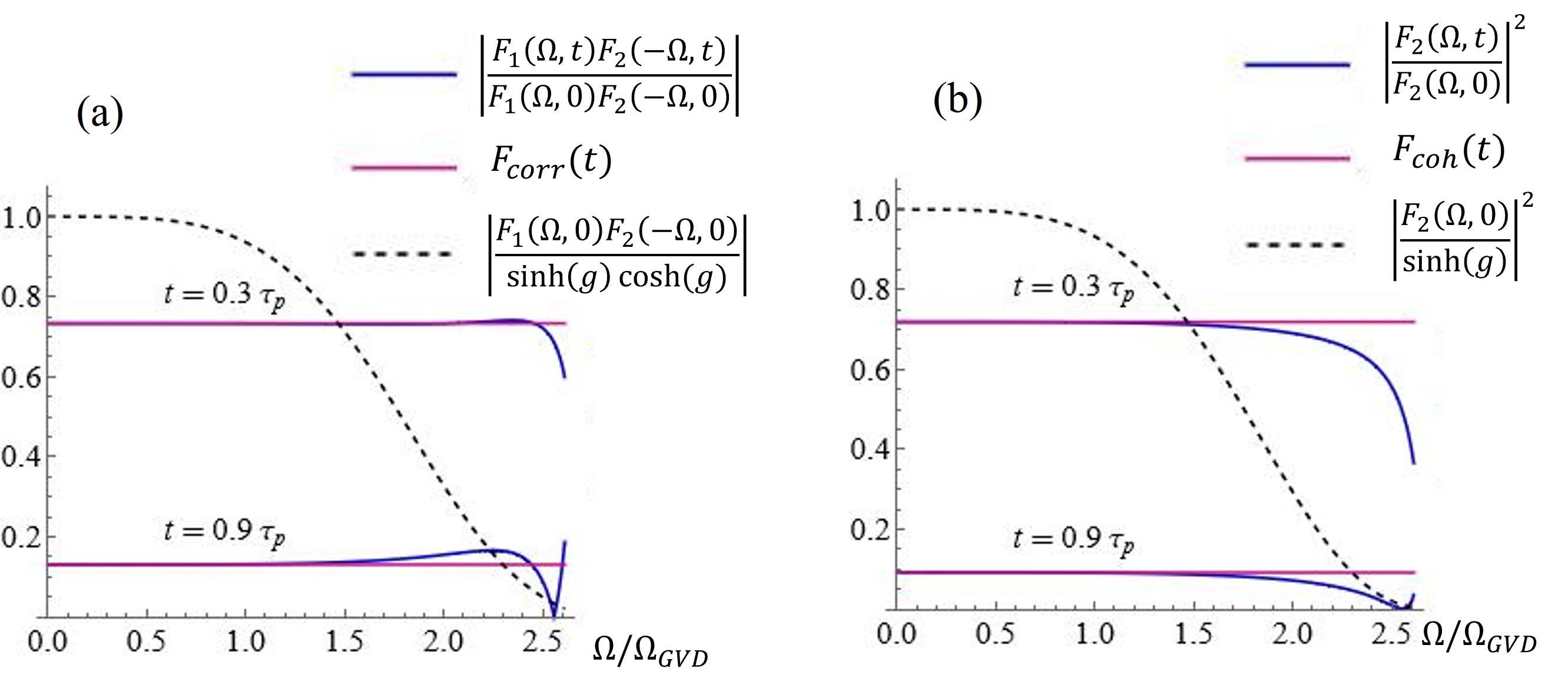}}}
 \caption{The blue solid lines plot the ratio  \eqref{FcorrS} as  functions of $\Omega$ in the central part of the spectrum at  $q=0$, and for $x=y=0$, for two values of the temporal coordinate, showing that they are almost costant, and equal to their approximated value $\Fcorr$ (a) or $\Fcoh$ (b)  (purple).   The dashed lines plot for comparison the corresponding spectral distributions (normalized to their peak value), 
  In this example  $g=1.8$, and other parameters  are  those of a 2mm BBO crystal, cut for collinear phase matching at $1030$nm, as in Fig. \ref{fig_bands}.}
 \label{fig_Kapprox1}
 \end{figure} 
 %%%%%%%%%%%%%%%%%%%%%%%%%%%%%%%%%%%%%%%
 We have tested the validity of this approximation  in several conditions: Fig. S4 reports an example of the ratios \eqref{FcorrS}, plotted as  functions of $\Omega$ in the central part of the spectrum at $q=0$, and for $x=y=0$, in the case of a BBO cut for collinear and degenerate phase matching (parameters as in Fig.\ref{fig_bands}).  When time (or equivalently the spatial coordinate) is chosen well inside the pump pulse profile, the approximation is almost perfect for $F_1 F_2$ (panel a) and slightly worse for $F_2^2$ (panel b). At larger times along the pump profile, the error made is more consistent, but obviously these points do not contribute  much because the fluorescence pulse has there a low intensity. \\
  %%%%%%%%%%%%%%%%%%%% Fig. S5 
 \begin{figure}[h]
 \centering
{{\includegraphics[width=0.96\linewidth]{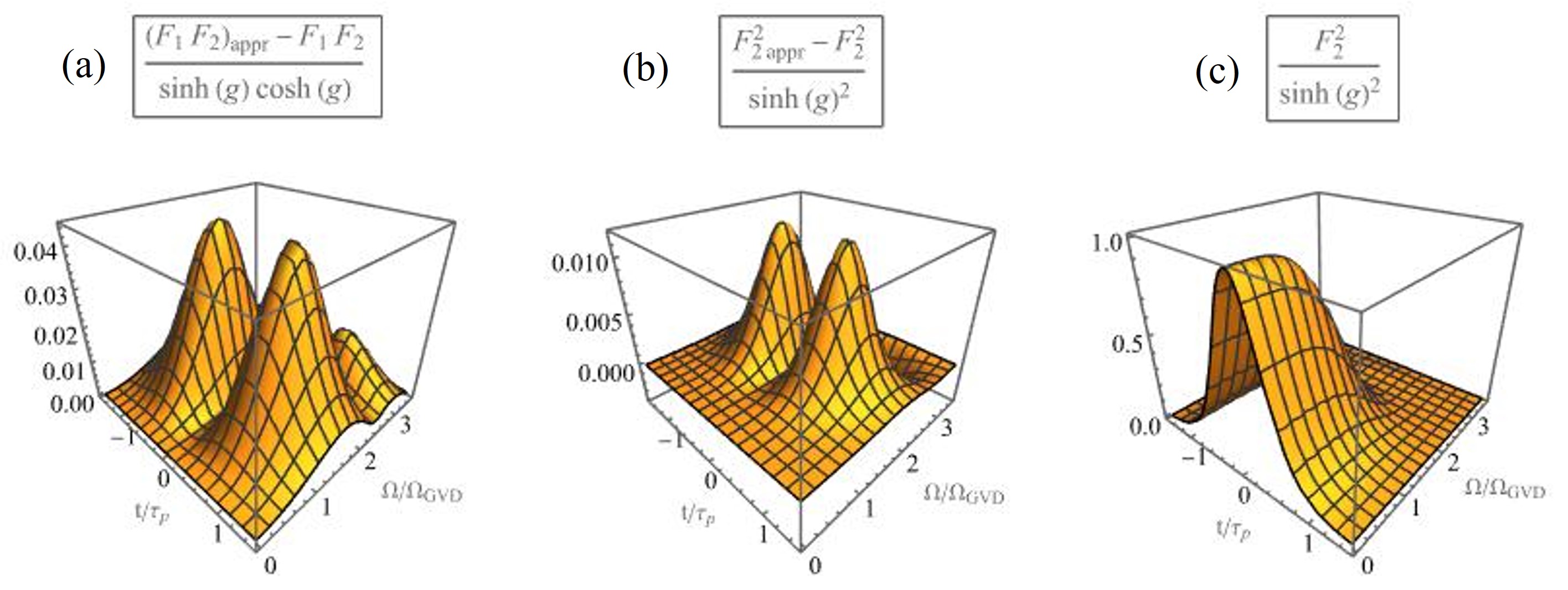}}}
 \caption{(a) and (b): Error made when substituiting the true kernels of the Boguliubov transformation\eqref{inout} with their factorized form defined by  Eq.\eqref{Fapprox}. (c) for comparison shows $F_2 (\w,\vxi)^2$. All the functions are plotted as functions of $\Omega$ and $t$ for $q=0$ and $x=y=0$, and are scaled to their peak values.  
   $g=1.8$, other parameters   as in Fig. \ref{fig_bands}.}
 \label{fig_Kapprox2}
 \end{figure} 
 %%%%%%%%%%%%%%%%%%%%%%%%%%%%%%%%%%%%%%%
 \noindent Fig. S5 reports directly the error made when one substitues 
  \beq
   \begin{aligned} 
& F_1 (\w, \vxi) F_2(-\w,\vxi)   \to \left( F_1 (\w, \vxi) F_2(-\w,\vxi) \right)_\text{appr}  =\Fcorr (\vxi) F_1(\w,0) F_2(-\w,0) , \\
 & F_2 (\w, \vxi)^2    \to  \left( F_2 (\w, \vxi)^2\right)_\text{appr} = \Fcoh (\vxi) F_2(\w,0)^2 
\label{Fapprox}
\end{aligned}
\eeq
 %%%%%%%%%%%%%%%%%%%% Fig. S6 
 \begin{figure}[h]
 \centering
{{\includegraphics[width=1.\linewidth]{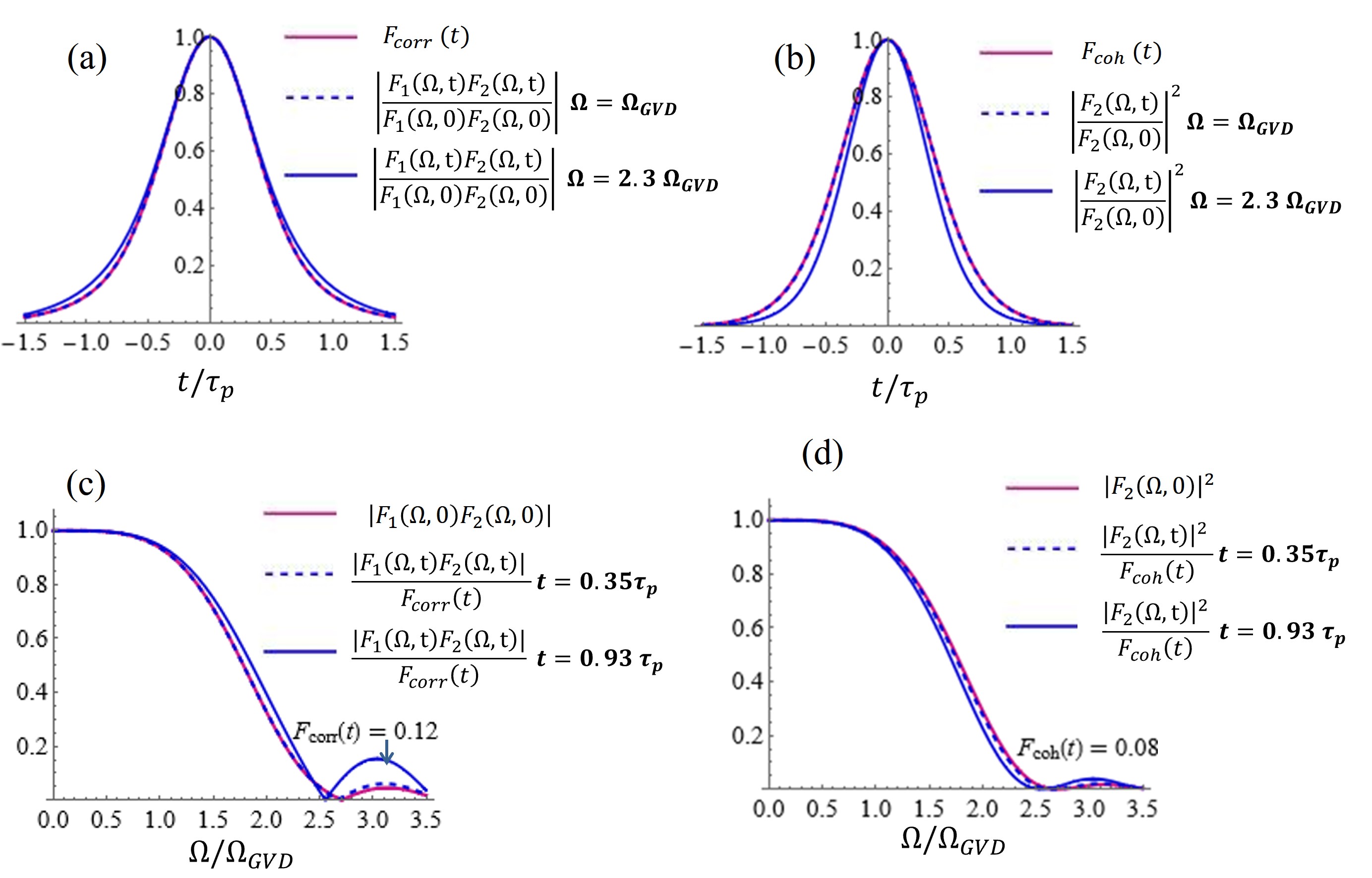}}}
 \caption{The plots  illustrates the validity of the approximation \eqref{FcorrS} in the space-time domain (a) and (b), and in the spectral domain (c) and (d), see text. 
 $g=1.8$, other parameters   as in Fig. \ref{fig_bands}.
 }
 \label{fig_Kapprox3}
 \end{figure} 
 %%%%%%%%%%%%%%%%%%%%%%%%%%%%%%%%%%%%%%%
 Fig. S6, finally,  illustrates the validity of the approximation in the space-time domain [panels (a) and (b)], and in the spectral domain [panels (c) and (d)].  (a) and (b)  compare  the temporal shape of the  distributions  $\Fcorr (0, t)$ and $\Fcoh (0,t)$  resulting from the factorized model \eqref{CorrSR} (purple lines), with the corresponding  ratios calculated in the model \eqref{Corr} at two different Fourier coordinates (dashed blue curve: $\Omega = \Omgvd, \, q=0$, solid blue curve $\Omega= 2.3 \Omgvd, \, q=0$), and shows again 
 that the approximation is excellent when the Fourier mode is chosen well inside the phase-matching bandwith, and become sligthly worse at the bandwidth borders. Panels (c) and (d) are the corresponding spectral conterparts, with red lines showing the approximated result of the factorized model of Eqs.\eqref{CorrSR} and the blue lines ploting instead the corresponding quantities calculated with the less approximated model \eqref{CorrSR}. 
 In these examples the gain is intermediate $g=1.8$, and we checked that the results appear definitely better  for low gain $g<1$, while we did not notice significant differences at higher gain. 
\par
With this approximation, the correlation and coherence functions take the factorized forms of   Eqs.\eqref{CorrSR}, which for example for the biphoton correlation $\Psi (\w,\wpr)$ is the product of a fast decaying correlation peak, function of the distance $\w_0= \wpr+\w$ and a slowly varying function of $\w$. This form is very advantageous for analytical calculations, but as already remarked,  when more precise quantitative evaluations are needed,  the less approximated form \eqref{Corr} of results can be instead used and numerically evaluated. 
%%%%%%%%%%%%%%%%%% FINE SUPPLEMENTARY %%%%%%%%%%%%

\end{document}